%% file: main.tex
\documentclass[acmsmall,manualscript,screen]{acmart}

\AtBeginDocument{%
  \providecommand\BibTeX{{%
      \normalfont B\kern-0.5em{\scshape i\kern-0.25em b}\kern-0.8em\TeX}}}

\input{macros}

\begin{document}
\title{A Survey of Smart Contract Formal Specification and Verification}

\author{Palina Tolmach}
\email{palina001@ntu.edu.sg}
\affiliation{
  \institution{Nanyang Technological University}
  \streetaddress{50 Nanyang Avenue}
  \country{Singapore}
  \postcode{639798}
}
\affiliation{
  \institution{Institute of High Performance Computing (A*STAR)}
  \streetaddress{1 Fusionopolis Way, \#16-16 Connexis North}
  \country{Singapore}
  \postcode{138632}
}

\author{Yi Li}

\author{Shang-Wei Lin}

\author{Yang Liu}
\email{{yi_li,shang-wei.lin,yangliu}@ntu.edu.sg}
\affiliation{
  \institution{Nanyang Technological University}
  \streetaddress{50 Nanyang Avenue}
  \country{Singapore}
  \postcode{639798}
}

\author{Zengxiang Li}
\email{zengxiang_li@outlook.com}
\affiliation{
  \institution{Institute of High Performance Computing (A*STAR)}
  \streetaddress{1 Fusionopolis Way, \#16-16 Connexis North}
  \country{Singapore}
  \postcode{138632}
}


\begin{abstract}
  A smart contract is a computer program which allows users to automate their actions
  on the blockchain platform.
  Given the significance of smart contracts in supporting important activities across industry
  sectors including supply chain, finance, legal and medical services, there is a strong demand for
  verification and validation techniques.
  Yet, the vast majority of smart contracts lack any kind of formal specification, which is
  essential for establishing their correctness.
  In this survey, we investigate formal models and specifications of smart contracts presented in the literature
  and present a systematic overview in order to understand the common trends. 
  We also discuss the current approaches used in verifying such property specifications and identify
  gaps with the hope to recognize promising directions for future work.
\end{abstract}




\thanks{This research is partially supported by the Ministry of Education, Singapore, under
its Academic Research Fund Tier 1 (Award No. 2018-T1-002-069) and Tier 2 (Award No. MOE2018-T2-1-068),
and by the National Research Foundation, Singapore, and the Energy Market Authority, under its Energy Programme
(EP Award No. NRF2017EWT-EP003-023). Any opinions, findings and conclusions or recommendations expressed in this material
are those of the authors and do not reflect the views of National Research Foundation, Singapore and the Energy Market Authority.}

\maketitle


\input{intro}
\input{formalisms}
\input{specifications}
\input{verification}
\input{conclusion}
\input{conclude}


\bibliographystyle{ACM-Reference-Format}
\bibliography{literature_short}


\end{document}
\endinput

%% file: macros.tex
\usepackage{xspace}
\usepackage{xcolor}
\usepackage{graphicx}
\usepackage[normalem]{ulem} 
\usepackage{amsmath}
\usepackage{enumitem}
\usepackage{microtype}
\usepackage{hyphenat}
\usepackage{zed-csp}

\usepackage{hyperref}
\usepackage[anythingbreaks]{breakurl}
\usepackage[capitalize]{cleveref}
\crefname{section}{Sect.}{Sects.}
\Crefname{section}{Section}{Sections}

\usepackage{booktabs}
\usepackage{multirow}
\usepackage{makecell}
\usepackage{ragged2e}
\usepackage{longtable}
\usepackage{rotating}
\usepackage{lscape}
\usepackage{graphics}

\usepackage{subfigure}
\usepackage{wrapfig}

\usepackage{listings}

\usepackage[many]{tcolorbox}
\usepackage{lipsum}

\usepackage{bussproofs}
\usepackage{stackengine}

\usepackage{pifont}
\usepackage{tikz}
\usetikzlibrary{calc}
\usetikzlibrary{shapes.geometric}
\usetikzlibrary{decorations.pathreplacing}
\usetikzlibrary{positioning}
\usetikzlibrary{calc}
\usetikzlibrary{arrows}
\usetikzlibrary{shadows}

\usepackage{datetime} 

\renewcommand\footnotetextcopyrightpermission[1]{}
\newcommand{\T}[1]{\rotatebox[origin=c]{90}{#1}}

\newcommand{\XSpace}[1]{}

\def\bitcoinA{%
  \leavevmode
  \vtop{\offinterlineskip 
    \setbox0=\hbox{B}%
    \setbox2=\hbox to\wd0{\hfil\hskip-.03em
    \vrule height .3ex width .15ex\hskip .08em
    \vrule height .3ex width .15ex\hfil}
    \vbox{\copy2\box0}\box2}}

\newcommand{\Ethereum}[1]{\textcolor{darkgray}{#1\textsuperscript{$\Xi$}}}
\newcommand{\Tezos}[1]{\textcolor{magenta}{#1\textsuperscript{\ding{118}}}}
\newcommand{\Hyperledger}[1]{\textcolor{red}{#1\textsuperscript{\ding{72}}}}
\newcommand{\EOS}[1]{\textcolor{violet}{#1\textsuperscript{\ding{61}}}}
\newcommand{\Bitcoin}[1]{\textcolor{blue}{#1\textsuperscript{\bitcoinA}}}
\newcommand{\Other}[1]{\textcolor{orange}{#1\textsuperscript{\ding{58}}}}

\newcommand{\TotalNumOfPapers}{202\xspace}
\newcommand{\ModelCheckCount}{18\xspace}
\newcommand{\TheoremProvCount}{20\xspace}
\newcommand{\ProgramVerCount}{42\xspace}
\newcommand{\SymbolicExCount}{22\xspace}
\newcommand{\RuntimeVerCount}{26\xspace}
\newcommand{\Query}{70\xspace}
\newcommand{\Snowballing}{85\xspace}
\newcommand{\Alert}{47\xspace}

\newcommand{\DefMacro}[2]{\expandafter\newcommand\csname rmk-#1\endcsname{#2}}
\newcommand{\UseMacro}[1]{\csname rmk-#1\endcsname}

\renewcommand{\paragraph}[1]{\vskip 0.05in \noindent\textit{#1.}}

\newcommand{\InputWithSpace}[1]{\bgroup\def\arraystretch{1.1}\input{#1}\egroup}
\newcommand{\Code}[1]{{\ifmmode{\mathtt{#1}}\else$\mathtt{#1}$\fi}}
\newcommand{\CodeIn}[1]{{\ifmmode{\mathtt{#1}}\else$\mathtt{#1}$\fi}}

\newcommand{\tool}[1]{\texttt{#1}\xspace}
\newcommand{\lang}[1]{#1\xspace}
\newcommand{\contract}[1]{#1\xspace}

\definecolor{silver}{RGB}{192,192,192}

\newcolumntype{R}[1]{>{\RaggedLeft\arraybackslash}p{#1}}
\newcolumntype{L}[1]{>{\RaggedRight\arraybackslash}p{#1}}
\newcolumntype{C}[1]{>{\centering\arraybackslash}p{#1}}

\newcolumntype{?}{!{\vrule width 1.3pt}}

\lstdefinestyle{tafile}{%
    basicstyle=\scriptsize\ttfamily,
    frame=single,
    rulecolor=\color{black},
    morekeywords={\$INHERIT, Update, Insert, Delete, reference, contain, call, HunkDep, Coverage},
    tabsize=2,
    numbers=left,                    
    numbersep=5pt,                   
    stepnumber=1,
    numberstyle=\tiny\color{gray},   
    keywordstyle=\sffamily,
}

\newtcolorbox{rqbox}[1]{%
    tikznode boxed title,
    enhanced,
    arc=0mm,
    boxrule=0.5pt,
    interior style={white},
    attach boxed title to top center= {yshift=-\tcboxedtitleheight/2},
    colbacktitle=white,coltitle=black,
    boxed title style={size=normal,colframe=white,boxrule=0pt},
    title=\textbf{Answer to }{\textbf{#1}}}

\hypersetup{final}

%% file: intro.tex
\section{Introduction}\label{sec:intro}

\begin{figure}[tb]
  \begin{minipage}{.42\linewidth}
    \includegraphics[width=1.01\textwidth]{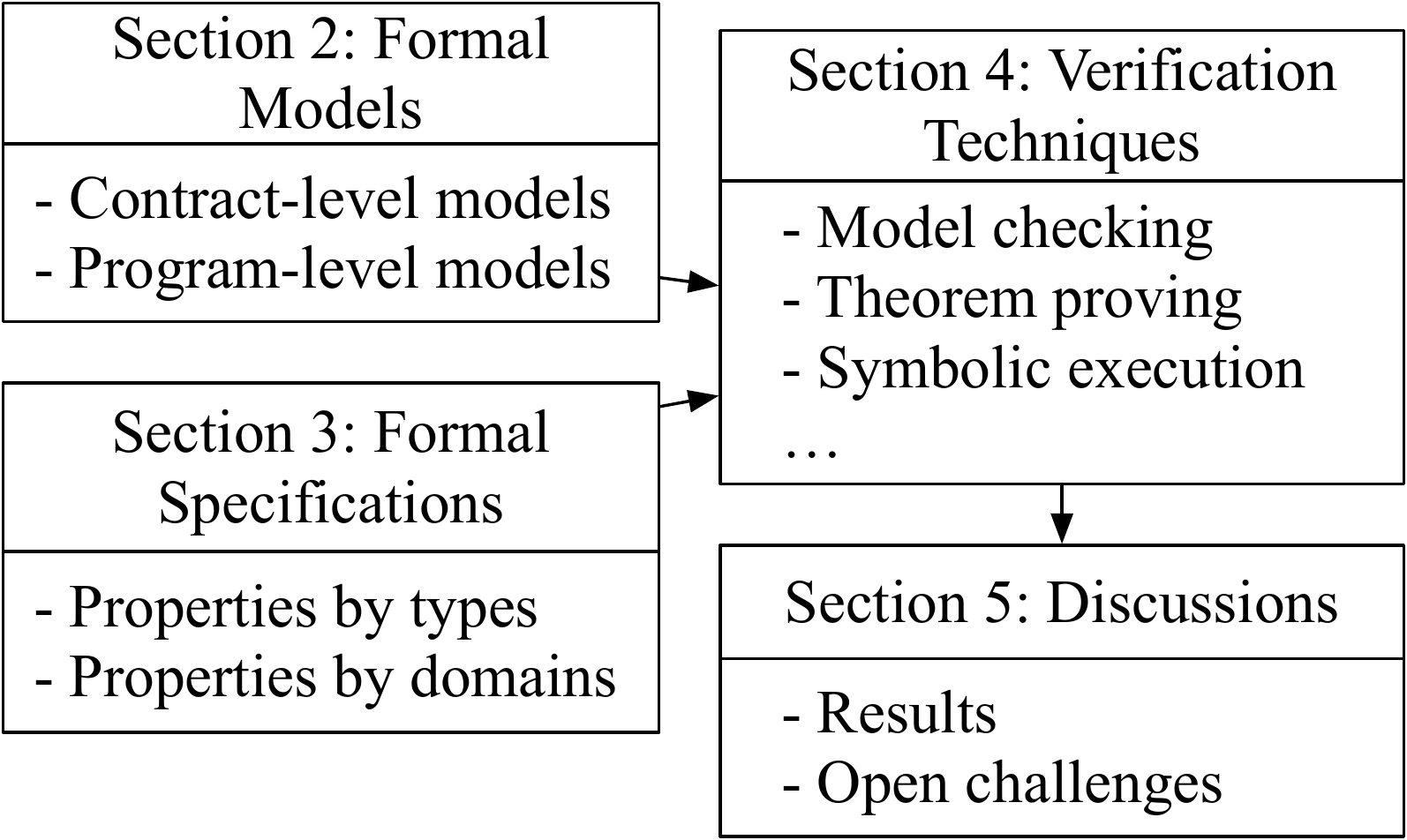}
    \caption{Organization of the paper.}\label{fig:organization}
  \end{minipage}
  \qquad
  \begin{minipage}{.5\linewidth}
    \footnotesize
\begin{verbatim}
("formal"
  ("verification" OR "modeling" OR "specification")
  "of smart contracts")
OR ("smart contract properties")
OR ("smart contract temporal properties")

("formal"
  ("verification" OR "modeling" OR "specification")
  "of smart contracts")
AND ("properties" OR ("temporal properties"))
\end{verbatim}%
\caption{Search queries used in Google Scholar.}\label{fig:queries}
\end{minipage}
\end{figure}

A proposal to store timestamped tamper-resistant information in cryptographically secure chained
blocks has gradually evolved from 1991~\cite{Haber1991}
into what we now know as \emph{blockchain}: a distributed
ledger shared between nodes of a peer-to-peer network following a certain consensus protocol,
and one of the fastest-growing technologies of the modern time.
The blockchain technology owes its initial popularity to Bitcoin~\cite{Nakamoto2008},
which appeared in 2008.
{Later, Ethereum~\cite{Wood2014} started the era of Blockchain 2.0 and expanded the capabilities and
applications of blockchain by introducing Turing-complete \emph{smart contracts}}.
The term \emph{smart contract} was originally proposed by Nick Szabo~\cite{Szabo1997},
who suggested encoding the contractual terms and protocols of the involved parties in the digital form.
In the blockchain context, smart contracts currently refer to computer programs executed on
top of blockchain, and we concentrate on this definition of a smart contract in this work.
In Ethereum, each contract is associated with an address, a balance, and a piece of executable code.
The functions in a contract can be invoked via transaction messages sent from platform users.
Most of the smart contracts are written in Turing-complete languages, such as
\lang{Solidity}~\cite{Solidity}.

The market cap of the Ethereum cryptocurrency was worth \$20 billion as of February 1, 2020, and
several hundred of thousands of transactions are processed each day~\cite{Etherscan}.
Unfortunately, the adoption of blockchain and smart contracts is also accompanied by severe
attacks, often due to domain-specific security pitfalls in the smart contract implementations.
Examples include theft of Ether through notorious attacks on the \contract{DAO}~\cite{DAO} and
\contract{Parity Multisig Wallet}~\cite{Parity2017Hacked},
as well as the freezing of users' funds in the same \contract{Parity Multisig Wallet}~\cite{Parity2017Frozen}
and the \contract{King of the Ether Throne}~\cite{KotET} smart contracts.
Extensive research has been done over the past several years---by technology giants, non-profit
organizations, and the research community---on the techniques for verifying smart contract
correctness.
In this survey, we review the recent advances in the applications of formal methods in the analysis
of smart contracts, with a focus on the domain-specific properties and various formalisms
supporting the specification and verification of such properties.

We organize the paper following a structure outlined in \cref{fig:organization}.
\emph{Formal verification} proves or disproves the correctness of a system by checking the
\emph{formal (mathematical) model} of the system against a certain \emph{formal specification}.
A specification is a set of \emph{properties} describing the desired behaviors of a smart contract,
usually defined by developers' intention.
Models and specifications can be defined at different levels of abstraction via various types of
formalisms, which we overview in \cref{sec:modeling,sec:specifications}, respectively.
\Cref{sec:specifications} additionally describes a taxonomy of smart contract properties, which
highlights the emerging trends in the literature.
\Cref{sec:verification} closes the loop by reviewing formal verification techniques
utilized to prove the correctness of smart contract models with respect to the desired properties.
Finally, \cref{sec:discussion} concludes the survey with our observations on major trends,
challenges, and future directions in the formal specification and verification of smart contracts.

\subsection{Methodology}\label{ssec:methodology}
\Cref{ssec:methodology,ssec:related-works} outline the
methodology that we followed to extensively collect the existing work related to smart contract
formal specification and verification, and an overview of the body of related works, respectively.

\paragraph{Research Scope}
This paper is aimed to overview, analyze, and classify approaches to formal
modeling of smart contracts, specification of smart contract properties, and techniques
employed in the verification of those.
The scope of the survey includes a study of common property specifications of smart contracts, characterizing their functional
correctness and security guarantees.
The focus of the survey is narrowed down to smart contract functional
behavior and does not embody problems related to other blockchain-related
execution aspects, such as scalability, consensus, interaction with
IoT systems, etc.
Therefore, the purpose of the paper is to provide answers to the following \emph{research
questions}:
\begin{itemize}[leftmargin=.36in,topsep=1pt]
\item[\textbf{RQ1:}] What are the formal techniques used for modeling, specification, and verification of
smart contracts? What are the common formal requirements specified and verified by these techniques?
\item[\textbf{RQ2:}] What are the challenges introduced by smart contract and blockchain
environment in formalizing and verifying smart contracts?
\item[\textbf{RQ3:}] What are the current limitations in smart contract formal
specification and verification and what research directions may be taken to overcome them?
\end{itemize}

\paragraph{Search Criteria}
In order to provide a complete survey covering most of the publications related to smart
contract specification and verification since 2014, we created a publication repository that
includes $\TotalNumOfPapers$ papers published from September, 2014 to June, 2020.
To collect the relevant literature for analysis, we have performed the search in Google Scholar,
ACM, IEEE, and Springer databases using the two queries shown in \cref{fig:queries}.

In October 2019, we have explored $10$ pages of the results retrieved for
each query and obtained $\Query$ papers that fall into the scope of the survey and
have been published since 2014.
Based on the collected set of papers, we performed snowballing and
inverse snowballing to obtain a more comprehensive view by collecting $\Snowballing$
more publications. Additionally, to provide an overview of the state-of-the-art,
we included another $\Alert$ recently published papers that were retrieved using
the two aforementioned queries from October, 2019 to June, 2020 by Google
Scholar at the time of writing. Overall, we collected and studied $\TotalNumOfPapers$
relevant papers.
The majority of them were published in the proceedings of 70 conferences and workshops
(including ACM CCS, ASE, ESEC/FSE, IEEE SANER, S\&P, IEEE/ACM ICSE, NDSS, ISoLA, FC, FM, and the
affiliated workshops) or one of 13 journals (including ACM TOSEM, IEEE TSE, and IEEE Access).
Using our search methodology, we additionally identified 53 articles that conduct
a survey on smart contract types, languages, vulnerabilities, and verification techniques,
or propose security and design patterns for their development.
We describe the collected articles in an online repository:
\url{https://ntu-srslab.github.io/smart-contract-publications/}.

In the presented work we do not claim to provide a complete set of properties that smart contracts
are supposed to meet. However, by means of the presented analysis and classification,
we seek to provide a firm formalized foundation and a stable structure for smart contract
property specifications, what we hope will facilitate the development of
correct, secure, and standardized smart contracts.

\subsection{Related Works}\label{ssec:related-works}
By now, a multitude of surveys addressing smart contract analysis has been published.
These include a review of security
vulnerabilities~\cite{Atzei2017,Yamashita2019,Groce2019,Sayeed2020,Mense2018},
verification approaches and
tools~\cite{Chen2019a,DiAngelo2019,Perez2019,Praitheeshan2019,Durieux2019,Feng2019,Murray2019,Chen2019,Ye2019},
formal specification and modeling techniques~\cite{Singh2019,Bartoletti2019a,Imeri2020}, and languages for
smart contract development~\cite{Parizi2018,Harz2018}.
We found several pieces of work discussing common design patterns of smart
contracts~\cite{Wohrer2018,Wohrer2018a,Xu2018,Worley2019},
as well as studies of smart contract platforms and practical
applications~\cite{Bartoletti2017,Wang2019,Atzei2019a,Wei2019}.
Another line of research focuses on revealing challenges and opportunities for safer smart contract
development~\cite{Magazzeni2017a,Zou2019,Zheng2020,Miller2018,Arias2019}.
In addition, some surveys analyze performance and scalability aspects of smart
contracts~\cite{Rouhani2019}.

The surveys most relevant to this work are concerned with the application of formal methods to
modeling smart contracts.
Singh et al.~\cite{Singh2019} analyzed 35 research works on formal verification and specification
techniques as well as languages for smart contracts, together with the issues and vulnerabilities
they address.
Bartoletti et al.~\cite{Bartoletti2019a} compared five formal modeling techniques for Bitcoin smart
contracts based on their expressiveness, usability, and suitability for verification.
Furthermore, Ladleif and Weske~\cite{Ladleif2019} proposed a framework for evaluating formal
modeling tools with respect to their capability to model a legal smart contract, and
assessed eight visual tools for smart contracts modeling using this framework.
Different from the discussed works, our survey proposes a coherent conceptual framework that helps
to establish the links between the approaches to formal modeling, specification, and verification.
Our analysis is also more comprehensive in terms of the types of formal techniques covered and
the application platforms considered.

There has been a body of literature dedicated to the collection of both bad and good ways of
writing \emph{secure} smart contracts.
On one hand, the common pitfalls discovered in past experiences should be documented and avoided in
the future development; and on the other hand, the best practices which are proven to work can be
summarized and adopted when facing a similar context.
Atzei et al.~\cite{Atzei2017} and some more recent
work~\cite{Yamashita2019,Groce2019,Sayeed2020,Mense2018} collected common vulnerability patterns
which might compromise the security of smart contracts.
Permenev et al.~\cite{Permenev2020} and Bernardi et al.~\cite{Bernardi2020} defined several classes
of properties for which security analysts should watch out when performing contract audit.
Several groups of researchers~\cite{Bartoletti2017,Wohrer2018,Wohrer2018a,Xu2018,Worley2019}
proposed design patterns for smart contracts, which are general and reusable solutions for specific
tasks.
In this survey, we look beyond the security aspect and aim to provide a general taxonomy of
domain-specific properties, which may also impact the correctness, privacy, efficiency, and
fairness of smart contracts.
We are also interested in establishing the links between the various types of properties and the
formalism used in supporting them.
We believe that this effort can help identify the gaps between the analysis needs and the existing
technology capability.

We observe a consensus from the previous surveys that specifications for smart contracts are of
paramount importance, but how to properly derive and represent them still remains an open challenge.
This impacts not only the development of safe smart contracts, but also their verification and
utilization~\cite{Grishchenko2018,Huang2019,Wang2019,Magazzeni2017a}.
According to a recent survey on smart contract developers by Zou et al.~\cite{Zou2019}, best
practices, standards, and provably safe re-usable libraries are listed among the most desired
improvements they would like to see to happen in the smart contract ecosystem.
While there are ongoing efforts towards this direction, such as the creation of
curated datasets of smart contracts~\cite{Durieux2019,Ghaleb2020Injection}, we aim to facilitate the adoption of formal specification and
verification by contributing the results of our analysis as well.

We adopt the same holistic view of verification techniques and tools for
smart contracts. Still, our categorization of formal verification approaches
is inspired by the previous studies on smart contract verification techniques
performed by Harz et al.~\cite{Harz2018}, Di Angelo et al.~\cite{DiAngelo2019},
and Miller et al.~\cite{Miller2018}.
In this survey, we make the following novel contributions.
\begin{itemize}[topsep=1pt,leftmargin=*]
  \item A comprehensive survey of the state of the art with broader scope;
  \item A four-layer framework for classifying smart contract analysis approaches;
  \item A taxonomy of smart contracts specifications in various domains;
  \item Discussions on observed trends and challenges in smart contract development and
  verification.
\end{itemize}

%% file: formalisms.tex
\section{A Taxonomy of Smart Contract Modeling Formalism}\label{sec:modeling}

This section presents the formalisms used for formal modeling of smart contracts
in the literature.
The formalization approaches for smart contracts can be classified into two broad categories:
\emph{contract-level} and \emph{program-level}, based on the level of abstraction at which modeling
and analysis is performed.
The \emph{contract-level} approaches are concerned with the high-level behavior of a smart contract
under analysis and usually do not consider technical details of its implementation and execution.
These approaches usually describe the interactions between smart contracts
and external agents mainly represented by users and the blockchain. The details of a smart
contract are typically abstracted into a set of public functions that are invoked by users and
observable results of their executions, such as emitted events or a change in a blockchain state.
The \emph{program-level} approaches mostly perform analysis on the contract implementation (i.e.,
source code or compiled bytecode) itself and thus are platform-dependent. By using program-level
representations, such approaches enable precise reasoning about low-level details of the smart contract
execution process.

\subsection{Contract-Level Models}\label{ssec:contract-model}
In the contract-level analyses, smart contracts are viewed as black boxes which accept transaction
messages coming from outside, and possibly perform some computation based on them.
This may result in externally observable events and even irreversible alteration to the blockchain
state.
The internal execution details and intermediate results are abstracted away in these types of
models.
Essentially, the contract-level models are concerned with the following concepts.
\begin{itemize}[topsep=1pt,leftmargin=*]
  \item \emph{Users} -- together with their balance,
  the transactions they initiate and the associated parameters such as the account addresses and
  the attached values. They can be any type of accounts which interact with the contract of interest---a user account
  or another contact. Some papers~\cite{Laneve2019,Atzei2019,Chatterjee2018} also consider the
  choices of transactions made by users under different circumstances (i.e., strategies),
  as well as knowledge that is available to users at a particular moment of a smart contract
  execution~\cite{Andrychowicz2014,Atzei2019,Meyden2019,Hirai2018}.
  \item \emph{Contracts} -- characterized by a set of publicly accessible functions
  and the externally visible effects of the function executions, e.g., changes of contract
  owners/balances, and emitted events/operations.
  As such, the behaviors of a contract are usually described by a set of
  \emph{traces}~\cite{Shishkin2018}, which are finite sequences of function invocations (i.e., the
  executed operations~\cite{Mavridou2019}), and properties are predicates on such traces.
  \item \emph{Blockchain State} -- including global variables referred to by the contracts and
  environment variables such as the timestamps and block numbers.
  More generally, blockchain state may also include the mining pool~\cite{Nehai2018,Abdellatif2018b} and the
  memory state of the contract execution environment~\cite{Mavridou2019}.
\end{itemize}
The contract-level models are effective in expressing properties regarding the interactions between
smart contracts and the external environment, and they are usually defined in terms of process
algebras, state-transition systems, and set-based methods.
The majority of such models are verified using techniques based on model
checking~\cite{Nehai2018,Mavridou2019,Alqahtani2020,Madl2019,Bai2018,Molina-Jimenez2018,Qu2018,Abdellatif2018b,Andrychowicz2014,Osterland2020}.

\subsubsection{Process Algebras}\label{sssec:process-algebras}
Process algebra~\cite{Baeten2005} is a family of mathematical approaches used to describe the
behaviors of distributed or parallel systems in the form of interacting concurrent processes.
The observable behaviors of a smart contract are similar to that of a shared-memory concurrent
program~\cite{Sergey2017} or a system with an interleaving mode~\cite{Qu2018}, which makes process
algebra a suitable candidate for high-level behavioral modeling.
Therefore, process algebra captures the interactions between concurrently acting users and one
or several interconnected smart contracts~\cite{Tolmach2021}.
For these interactions, correctness can be assured through a translation of a smart
contract to one of the process-algebraic formalisms. An alternative approach involves
implementation of a smart contract in a process-algebraic domain-specific language (DSL).

\begin{figure}[tb]
    \begin{minipage}{0.5\textwidth}
        \includegraphics[width=\textwidth]{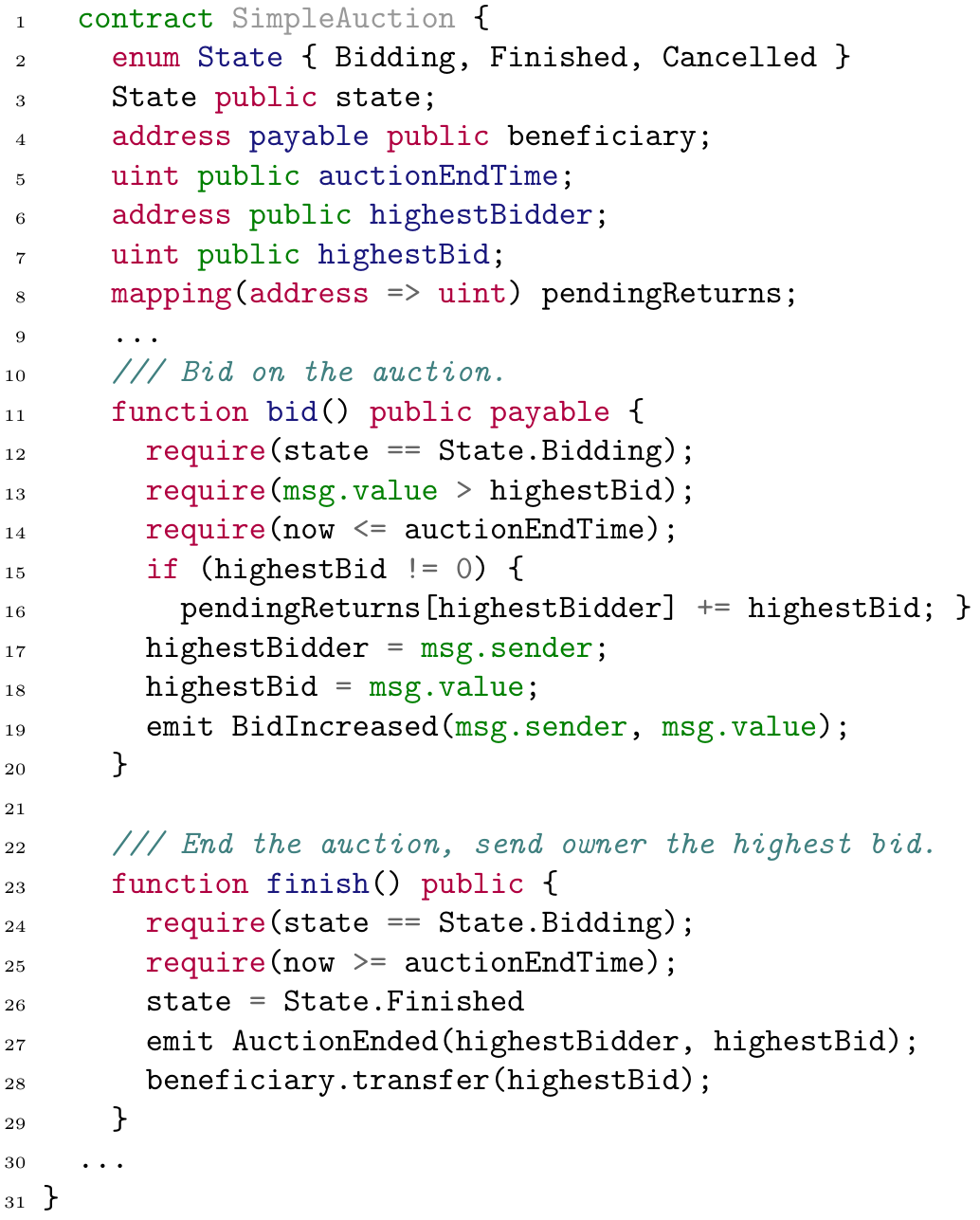}
        \caption{Solidity source code of Simple Auction smart contract adapted from~\cite{SolidityExamples,Mavridou2018}.}\label{fig:auction_code}
      \end{minipage}
    \quad
    \begin{minipage}{.4\textwidth}
    \scriptsize
    \begin{flalign*}
        & //\text{@@} SimpleAuction\text{@@} &\\
        &enum\{Bidding,Finished,Cancelled\}; &\\
        &var\: state = 0; &\\
        &var\: beneficiary = 0; &\\
        &var\: auctionEndTime = 0; &\\
        &var\: hBidder = 0; var \: hBid = 0; &\\
        &var\: pReturns[2^{256}-1]; &\\
        & \dots &\\
        &\mathbf{Bid}(msg,blk) = \mathbf{if} (state == Bidding\: \&\& \: msg.value > Bid &\\
        &\quad \&\& \: blk.timestamp <= auctionEndTime) \{ &\\
        &\qquad \mathbf{if} (hBid != 0) \{ &\\
        &\qquad\quad pReturns[hBidder] = pReturns[hBidder] + hBid;\mkern-9mu\} &\\
        &\qquad updateBid\{hBid = msg.value; &\\
        &\qquad\quad hBidder = msg.sender;\mkern-9mu\} \rightarrow hBidIncreased \rightarrow Skip\}; &\\
        \\
        &\mathbf{Finish}(blk) = \mathbf{if} (state == Bidding\: \&\& &\\
        &\quad blk.timestamp >= auctionEndTime) \{ &\\
        &\qquad endAuction\{state = Finished;\mkern-9mu\} \rightarrow &\\
        &\qquad transfer.beneficiary.hBid \rightarrow auctionEnded \rightarrow Skip\}; &\\
        &\cdots
      \end{flalign*}
      \caption{A CSP$\#$ model of Simple Auction smart contract.}\label{fig:auction_csp}
    \end{minipage}
  \end{figure}

Given that users interact with an Ethereum contract by calling its public functions, Qu et
al.~\cite{Qu2018} and Li et al.~\cite{Li2019bnb} translate such functions from Solidity to process
notations defined in \emph{Communicating Sequential Processes}~(CSP) and a variant of an applied
$\pi$-calculus \emph{SAPIC}, respectively.
In case transactions are not atomic, the interplay between the functions invoked by different users may lead to
risk conditions, which are manifested by vulnerable sequences appeared in the execution trace~\cite{Qu2018}.
The authors of~\cite{Li2019bnb} identify violations of trace-level safety properties that are
commonly analyzed for token implementations, such as the immutability of the total number of tokens
in a smart contract.
As an example, \cref{fig:auction_code,fig:auction_csp} demonstrate partial source code
and a CSP model of the \contract{Simple Auction} smart contract, which is based on
an example from the Solidity documentation~\cite{SolidityExamples}
and the \contract{Blind Auction} smart contract provided in~\cite{Mavridou2018}.
Within the contract, each user is allowed to place his bid until the deadline
is reached. After the auction is finished, the highest bid is transferred to the
beneficiary, while the unsuccessful bidders can withdraw their bids.
Alternatively, the auction can be canceled at the bidding stage. Since the users participate
in the auction by invoking functions of a smart contract, \Cref{fig:auction_code} shows the Solidity
implementation of two functions, namely, \texttt{bid()} and \texttt{finish()}.
Following the approach adopted by the authors of~\cite{Li2019bnb,Qu2018},
\cref{fig:auction_csp} represents each of these functions as a process: \texttt{Bid(msg,blk)} and
\texttt{Finish(blk)}, respectively. The processes are specified in CSP$\#$---a CSP-based formal language
of the PAT tool~\cite{Sun2009PAT}, which supports both process and programming constructs.
Each process describes the corresponding actions of a smart contract and its users
in terms of CSP \emph{events} as well as changes in a smart contract state implemented via shared variables.
Being a high-level representation, the model abstracts low-level execution
details, including that of the \texttt{transfer()} function, with the corresponding
events. Block and transaction properties are represented using the \texttt{msg}
and \texttt{blk} process parameters, similar to the globally available variables
in Solidity. For brevity, the model of user behavior is omitted.

A process-algebraic model of a smart contract is adopted by several domain-specific programming
languages.~\cite{Bartoletti2019,RhoLang}.
\emph{Rholang}~\cite{RhoLang} is a Turing-complete language for the RChain platform, which supports concurrency and is
based on the reflective higher-order process calculus $\rho$-calculus, a variant of
$\pi$-calculus. Smart contracts in \lang{Rholang} are processes that are triggered
by messages from agents: users or other contracts.
\emph{BitML}~\cite{Bartoletti2019} is a high-level smart-contract language
for Bitcoin,   which is based on a process calculus with symbolic semantics.
It is showcased in a series of publications by Bartoletti et
al.~\cite{Atzei2019,Bartoletti2019} how security properties, as well as arbitrary
temporal properties, can be expressed and checked in \lang{BitML}.
To ensure correctness of a smart contract under different scenarios, \lang{BitML} also
supports specification of user strategies.
Furthermore, smart contracts can also be soundly translated from \lang{BitML} into transactions---a
Bitcoin representation of a smart contract.
The behavior of contract users can also be partially defined in a Turing-complete process
calculus \emph{scl} proposed by Laneve et al.~\cite{Laneve2019}.
\lang{scl} models behaviors of a smart contract and a user as a parallel composition, where
a user acts nondeterministically, while the behavior of a smart contract is determined by its
execution logic.
The \lang{scl} models are in fact transition systems that are summarized by a first-order logic
formula describing the values of the objective function for all possible runs and user strategies.
Then, the game-theoretical analysis is performed to identify strategies that help users maximize
their profits or minimize losses.

  \subsubsection{State-Transition Systems}\label{sssec:transition-systems}
The behaviors of a smart contract can be naturally interpreted as a state-transition system, and in
fact, Solidity encourages contracts being modeled as state machines~\cite{StateMachine}.
Some next-generation smart contract languages, such as \lang{Obsidian}~\cite{Coblenz2019obsidian} and
\lang{Bamboo}~\cite{Bamboo}, follow a state-oriented programming paradigm which makes transitions
between states explicit and therefore provides better security guarantees.
When it comes to verification, depending on the properties interested, there are also a wide range
of choices in modeling smart contracts as state-transition systems.
For instance, timed automata, Markov decision processes, and Petri nets are used to capture the
time, probabilistic, and multi-agent interactive properties of the systems, respectively.
State-transition models are typically verified by model checkers against contract-specific
properties, in addition to generic temporal logic properties such as deadlock-/livelock-freedom
(C.f.~\cref{ssec:model-checking}).
\Cref{sssec:properties-security} provides a discussion on how the violation of these properties
can lead to serious bugs.
A variety of analysis frameworks already exist for the validation and verification of
state-transition systems, e.g., \tool{BIP} and \tool{cpntools}, etc.
Such verification capacities combined with the ability to visually define a state-transition
system make this formalism useful for smart contract
synthesis~\cite{Mavridou2018,Mavridou2019,Zupan2020}.

\emph{Petri net} (PN) is a mathematical language used to describe behaviors of concurrent systems.
Unlike the aforementioned formalisms, it provides a graphical notation which portrays a
system using a set of alternating places and transitions.
PN tools allow a user to visually specify a smart contract model and simulate its possible
behaviors.
Similar capabilities of the \emph{Colored Petri net}~(CPN) extension and the supporting
\tool{cpntools} toolkit are utilized for security analysis of a Solidity
contract~\cite{Liu2019a,Duo2020}.
These two techniques construct CPN models of a contract and its users---manually~\cite{Liu2019a}
or automatically~\cite{Duo2020}---to identify if a vulnerable state of a net is reachable.
The authors of~\cite{Duo2020} make the simulation more realistic by additionally modeling the
execution process of EVM, considering gas consumption as well.
The ability to visually define and execute a PN model also helps in synthesizing smart contracts
from user inputs.
A framework by Zupan et al.~\cite{Zupan2020} generates a Solidity smart contract from a model
defined as a PN workflow---a subtype of a PN describing business processes.
Similar to the discussed PN approaches~\cite{Liu2019a,Duo2020}, Zupan et al. automatically validate
the model according to typical properties such as the absence of transitions that can never be
executed~\cite{Duo2020,Liu2019a,Zupan2020} or reachability of a final state~\cite{Zupan2020}.
Through the visualization and validation of the model and its execution, a correct-by-design smart
contract can be generated by a user without prior knowledge of smart-contract programming.
The authors expect the technique to facilitate the application of smart contracts in use cases
including supply chain~\cite{Zupan2020} and legal contracts~\cite{Ladleif2019}.

For a similar purpose, several frameworks synthesize smart contracts from a specification
defined using industry standards for visual modeling, such as \emph{Business Process Model and Notation}~(BPMN)
and \emph{Unified Modeling Language} (UML). \tool{Caterpillar}~\cite{Lopez2018, DiCiccio2019}
and \tool{Lorikeet}~\cite{DiCiccio2019} translate BPMN models, which describe collaborative business processes such as supply chain, to
Solidity and deploy them to blockchain.
A Solidity implementation of a cyber-physical system is generated from a UML state diagram
in~\cite{Garamvolgyi2018}.
BPMN and UML notations provide the capabilities to specify proper access control~\cite{Garamvolgyi2018,DiCiccio2019},
control flow~\cite{Lopez2018}, and asset management~\cite{DiCiccio2019} in the generated
contracts. Still, despite being widely used in the industry, BPMN and UML lack well-established formal
semantics, which complicates the analysis of such models. A possible way to validate
a BPMN model involves translation to a formal language, such as PN, as performed by
Garcia et al.~\cite{Garcia2017}.

\begin{figure}[tb]
    \includegraphics[width=.5\linewidth]{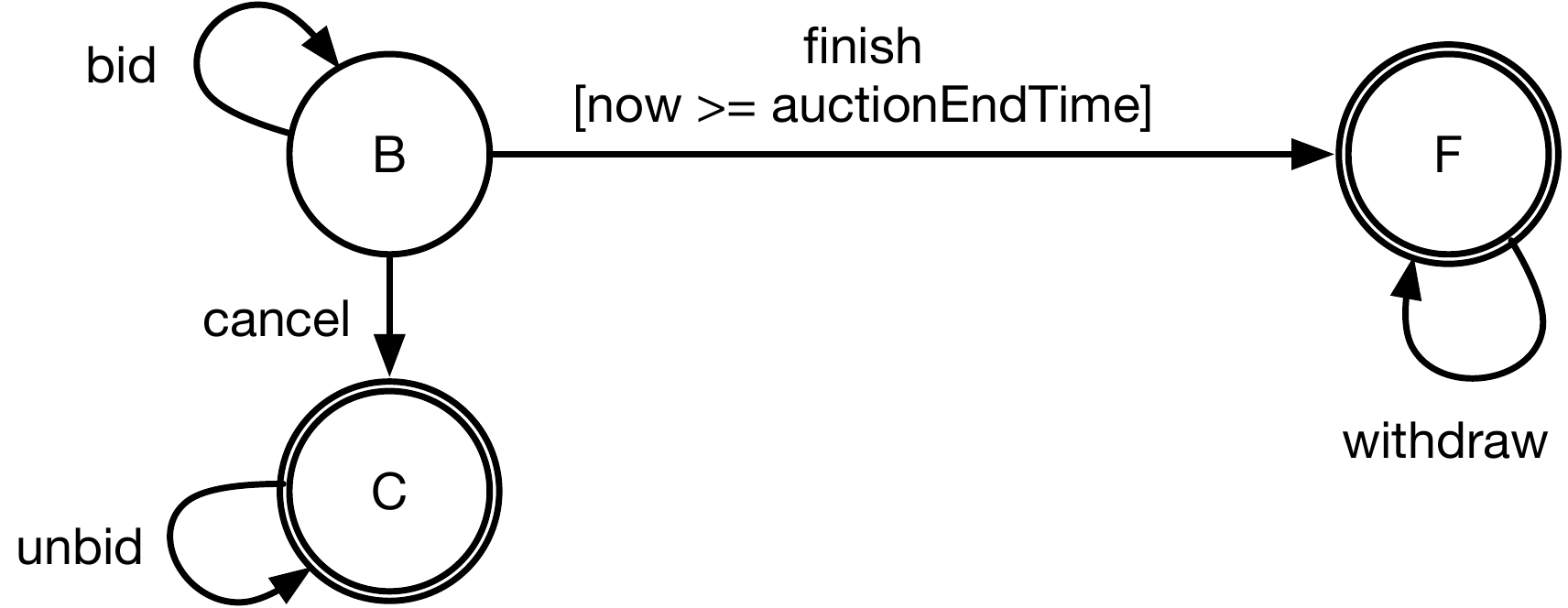}
    \caption{A state-transition model of \contract{Simple Auction} smart
    contract~(\cref{fig:auction_code}) adapted from~\cite{Mavridou2018}.}\label{fig:auction_fsm}
\end{figure}

\emph{Behavior-Interaction-Priority}~(BIP) is a layered framework used to model the interaction
between smart contracts and users. BIP models of smart contracts usually consist of two layers:
the components of the \emph{Behavior} layer describe the logic of each smart contract as an FSM,
while the \emph{Interaction} layer defines the principles of their
communication~\cite{Nelaturu2020,Alqahtani2020}.
Translation of an FSM model, visually defined in the \tool{FSolidM} framework~\cite{Mavridou2018},
to BIP and, then, an input language of a \tool{NuSMV} model checker, allows \tool{VeriSolid}~\cite{Mavridou2019}
to generate provably safe smart contracts (in contrast to UML- and BPMN-based
frameworks discussed earlier~\cite{Lopez2018, DiCiccio2019,Garamvolgyi2018}).
\Cref{fig:auction_fsm} presents an FSM model of the \contract{Simple Auction} smart contract~(\cref{fig:auction_code}),
inspired by a model of the \contract{Blind Auction} contract defined using \tool{FSolidM}~\cite{Mavridou2018}
and \tool{VeriSolid}~\cite{Mavridou2019}. The model in~\cref{fig:auction_fsm} depicts possible transitions between
three smart contract states: \texttt{Bidding} (B), \texttt{Finished} (F), and \texttt{Canceled} (C).

\emph{Timed automata} (TA) capture the temporal behavior of a system. TA models allow
measuring the impact of time on the execution of a smart contract~\cite{Abdellatif2018b}
or reasoning about time constraints, which are essential in legal and commitment
smart contracts~\cite{Clack2018}. Different from other approaches based on
BIP~\cite{Nelaturu2020,Alqahtani2020,Mavridou2019}, Abdellatif et al.~\cite{Abdellatif2018b}
build TA models of not only a contract, but also its users, and a simplified blockchain
mechanism, including mining of transactions in blocks. The interactions between these components
are analyzed by statistical model checking to evaluate the chances of a successful attack on
the contract. Andrychowicz et al.~\cite{Andrychowicz2014} use TA to describe behaviors of
parties in a Bitcoin timed commitment. Commitment protocols require participants to
exchange cryptographic hashes of the secret values they commit to before a stipulated time. To analyze correctness of this protocol,
the authors additionally equip each TA with a structure describing the party's knowledge.
The blockchain structure is maintained by a special agent, who is also modeled as an automaton.

A way to reason about the nondeterministic execution of a smart
contract involves modeling it through probabilistic and multi-agent extensions of transition systems, such as
\emph{Markov decision processes} or \emph{strategic} and \emph{concurrent games}~\cite{Bigi2015,Chatterjee2018,Laneve2019,Meyden2019}.
To account for all possible user behaviors, Meyden~\cite{Meyden2019}
regards an atomic swap smart contract as a concurrent game. Both a model and its properties
refer to the strategies of the contract participant. The authors apply the \tool{MCK} model checker
to establish correctness and fairness of a contract under the specified user behaviors.
Chatterjee et al.~\cite{Chatterjee2018} pursue a similar goal, however, their analysis
of a concurrent game model is based on its objective function, similar to the \emph{scl}
process-algebraic technique~\cite{Laneve2019}. To certify fairness and correctness of
a smart contract, the authors of~\cite{Chatterjee2018} examine the expected payoff of
its users and potential incentives for dishonest behavior.
Different from the game-theoretical techniques, Madl et al.~\cite{Madl2019} capture the interaction
between the users of a marketplace contract by describing their behaviors as \emph{interface automata}.
The authors verify a composition of the automata in a model checker to ensure
that participants can successfully cooperate.

\subsubsection{Set-Based Methods}\label{sssec:set-based}
Set-based modeling frameworks \emph{Event-B} and \emph{TLA+} utilize set theory and logic to
formally specify systems, and have been applied on smart contracts as well.
Analogically to other formal languages, such as PN, set-based frameworks
support the tools for model analysis, making them useful in the implementation of correct-by-design
smart contracts~\cite{Banach2020,Xu2019} and verification of the existing ones~\cite{Zhu2020}.
The authors of the identified publications~\cite{Zhu2020,Banach2020,Xu2019}
utilize set-based formalisms to reason about safety properties of smart contracts.
Although TLA+ supports temporal logic for specifying liveness, this feature has not yet been used
to formalize smart contract requirements.

Models written in Event-B are discrete transition systems which were demonstrated
to be a suitable formal representation of a smart contract. Zhu et al.~\cite{Zhu2020}
automatically translate a contract written in a subset of Solidity to
an Event-B model for refinement and verification via the \tool{Rodin} platform.
TLA+ is a formal specification language for concurrent systems. That allows
Xu et al.~\cite{Xu2019} to model behaviors of interacting participants of a legal smart
contract.
Although the authors of~\cite{Xu2019} do not translate their TLA+ specification
to executable code, they propose a set of design patterns to protect them from
common security vulnerabilities of smart contracts.
In general, the set-based correctness requirements for smart contracts specify
values allowed to be taken by contract variables~\cite{Banach2020,Zhu2020},
possible states of contract participants~\cite{Xu2019}, or the users that are
authorized to invoke contract functions~\cite{Zhu2020}.
Xu et al.~\cite{Xu2019} also verify whether events representing the actions of contract
parties are permitted according to the legal contract.

\subsection{Program-Level Models}\label{ssec:program-model}
The contract-level models are useful in reasoning about high-level interactions between the
contracts and external parties, but do not help much in understanding the internal details
about the contract execution.
The latter is essential in establishing properties concerning the correctness and security of
smart contracts.
The program-level models aim to provide a white-box view of the target contracts based on
lower-level representations, such as the source code, compiled bytecode, as well as derived
analysis artifacts including AST, data- and control-flow graphs.

The following are concepts that are relevant to this level of abstraction.
\begin{itemize}[topsep=1pt,leftmargin=*]
  \item \emph{Abstract Syntax Tree} (AST) -- representing smart contract source code (e.g., in
  Solidity) as hierarchical tree structures. It is often used to perform lightweight syntactic
  analyses on the contract implementations.
  \item \emph{Bytecode}~\cite{Permenev2020} (or \emph{opcode}) -- written in low-level machine
  instructions, which are closely tied to the execution environments (e.g., Ethereum Virtual
  Machine (EVM)). Since bytecode is obtained after compilation, it better reflects the
  machine-level specifics (e.g., gas consumption and exceptions), but at the same time may lose
  important source-level information.
  \item \emph{Control Flow Graph} (CFG) -- a graph representation of all program paths which might
  be traversed during execution. The CFG can be obtained from bytecode and is often used in static
  analyses of smart contracts, such as symbolic execution and automated verification.
  \item \emph{Program Traces} -- sequences of instructions (usually in bytecode) and events
  collected from executions at runtime. These traces reflect the exact behaviors of the contracts
  under specific inputs, which can be used as a source for performing dynamic analyses and runtime
  verification.
\end{itemize}
The program-level models preserve low-level execution details, and are, therefore, widely used in
finding vulnerabilities and checking other security-related properties.

\subsubsection{AST-Level Analyses}\label{sssec:ast}
Smart contract analyses performed directly on ASTs or similar parse-tree structures are often
aimed at the checking of predefined code patterns~\cite{Yamashita2019,Tikhomirov2018,Lu2019}.
For example, Yamashita et al.~\cite{Yamashita2019} scan ASTs of Hyperledger Fabric~\cite{Hyperledger} contracts
written in Go to find dangerous code patterns and library imports.
Another AST-based analysis is successfully applied in checking the compliance of a smart contract
with respect to the ERC20~\cite{ERC20} standard by Chen et al.~\cite{Chen2020}.
Moreover, AST can be used to facilitate lightweight pre-analyses: e.g., locating arithmetic
operations that may cause overflows or assertions~\cite{Viglianisi2020}, decoding the memory layout
information to allow deeper inspection~\cite{Bragagnolo2018}, and performing systematic
instrumentation, such as vulnerability checks, to the contract code~\cite{Akca2019,Samreen2020}.
Such pre-analyses can help improve the effectiveness and scalability of the more heavyweight
downstream analyses.
For instance, AST is the natural abstraction of smart contract code when it comes to
statistical~\cite{Liu2018} and deep learning-based analyses~\cite{Huang2019Update,Gao2019clone}.

The obvious drawback of these purely syntactic techniques is that they do not necessarily respect
the operation semantics or the execution environment of smart contracts, thereby compromising the
soundness/completeness of the analyses.
For example, the gas consumption plays an important role in the functional correctness and security
of Ethereum smart contracts, but is often neglected~\cite{Tikhomirov2018,Lu2019}.
To allow for more sophisticated analyses, ASTs are often translated to derive other intermediate
representations, such as CFG~\cite{Akca2019,Feist2019}, inheritance graph~\cite{Feist2019}, and
inter-contract call graphs~\cite{Chen2020}.

\subsubsection{Control-Flow Automata}\label{sssec:cfg}
\emph{Control-flow graph} (CFG) is often used to describe the operational semantics of a program.
As the name suggests, CFG is a labeled directed graph where the graph nodes correspond to program
locations and the edges correspond to possible transitions between the program locations.
Thus, CFG can also be viewed as a type of state-transition system, except that it more closely
reflects contract program executions than the contract-level state transitions discussed in
\cref{ssec:contract-model}.
CFGs of smart contracts are mostly constructed from the compiled code---either EVM bytecode of
Ethereum contracts~\cite{Luu2016} or WebAssembly (WASM) bytecode of EOSIO~\cite{EOS}
contracts~\cite{He2020,Quan2019}.
Still, recovering CFGs from low-level bytecode is non-trivial, at least in the case of Ethereum.
There are several reasons for this.
First, EVM is a stack-based machine, therefore, recovering the target address of a jumping
instruction (destination of an edge) calls for context-sensitive static analysis capable of
tracking the state of stack~\cite{Chang2018,Albert2020CFG}.
Furthermore, smart contract execution comprises a sequence of function invocations and, often,
calls to other accounts, which demands inter-procedural and inter-contract analyses, respectively.

The types of analyses run on CFGs include control-/data-flow analyses and symbolic execution,
which have matured tool supports for traditional software programs.
Some of these tools and techniques are applied directly on the CFGs extracted from smart contracts,
while others are custom built to accommodate particular language features of contract programs.
As one example, \tool{Ethainter}~\cite{Brent2020} performs information flow analysis to reveal
composite vulnerabilities, i.e., reached only through a sequence of transactions, from traces of
Solidity contracts.
In order to reuse the existing program analysis tool-chain, namely, Datalog and Souffl{\'e}, the
EVM bytecode is decompiled using \tool{GigaHorse}~\cite{Grech2019} to recover the CFG with data-
and control-flow dependencies.
The same tool-chain is also used by \tool{Securify}~\cite{Tsankov2018} and
\tool{Vandal}~\cite{Brent2018} in vulnerability detection after CFGs are constructed.
In contrast, many custom symbolic execution engines have been built to traverse CFGs and check for
interesting properties along the way.
Some of the most notable ones include \tool{Oyente}~\cite{Luu2016},
\tool{Mythril}~\cite{Mueller2018}, and \tool{Manticore}~\cite{Mossberg2019}.
These approaches are designed to identify vulnerable code
patterns~\cite{Luu2016,Mueller2018,Mossberg2019}, integer arithmetic bugs~\cite{Torres2018},
honeypots~\cite{Torres2019}, gas-inefficient bytecode patterns~\cite{Chen2017}, etc.

To allow specification and verification of custom properties, several tools support extensions to
CFGs with annotations.
\tool{Annotary}~\cite{Weiss2019} extends \tool{Mythril}~\cite{Mueller2018} to also capture
inter-contract and inter-transaction control-flows.
\tool{Solar}~\cite{Li2018} allows annotations to a smart contract with constraints and invariants.
\tool{SmartScopy}~\cite{Feng2019a} supports vulnerability patterns to be defined in their query
language based on Racket.

\subsubsection{Program Logics}\label{sssec:program-logic}
To rigorously reason about the correctness of programs, various forms of program logic have been
developed over the years, and some of them are successfully applied on smart contacts.
A \emph{program logic} is a proof system with a set of formal rules that allow to statically
reason about behavior of a program~\cite{Appel2014}.
For example, Hoare logic~\cite{Amani2018}, rewriting logic, and reachability logic~\cite{Park2018}
were used to prove correctness conditions (C.f.~\cref{sssec:program-specification}) for smart
contracts. An epistemic logic~\cite{Hirai2018} describes the interactions
between smart contract participants in terms of their knowledge, which facilitates verification
of commitment and swap protocols (C.f.~\cref{sssec:properties-privacy}).
A combination of authority algebra and FOL enables reasoning about trust and accountability in a smart
contract~\cite{Dargaye2018}.
Correctness of smart contracts is also verified through a dynamic logic developed
for Java programs~\cite{Ahrendt2019a,Beckert2018,Beckert2019}.
Finally, formalisms that are based on deontic logic enable specification and
verification of legal smart contracts, which are difficult to define formally
otherwise~\cite{Ladleif2019}.

Hoare and reachability logics assist reasoning about smart contract correctness
by different formal techniques.
Based on a formalization of EVM in \emph{Lem}~\cite{Hirai2017},
Amani et al.~\cite{Amani2018} and Duo et al.~\cite{Duo2020} defined a sound Hoare-style
program logic of Ethereum contracts. In verification of properties, however,
the former approach relies on theorem proving, while the latter translates
a formal model of a contract to CPN for model checking.
In other approaches based on theorem proving, Hoare-style properties
are verified for smart contracts written in \lang{Michelson}~\cite{Bernardo2019}---a smart contract
language of the Tezos~\cite{Tezos} platform, or \lang{Lolisa}~\cite{Yang2019a}---a formal
syntax and semantics of a Solidity language subset.
Apart from the tools based on Hoare logic, $\mathbb{K}$ is a framework for design and
development of language semantics based on rewriting logic~\cite{Park2018}.
$\mathbb{K}$ has been used to define a complete executable formal semantics
of EVM~\cite{Park2018} and Solidity~\cite{Jiao2020}, and a formal specification
of an intermediate LLVM-like language for smart contracts \lang{IELE}~\cite{Kasampalis2019}.
Properties that can be automatically proved by $\mathbb{K}$ include pre- and postconditions as
well as reachability claims that are verified through matching logic.
The expressiveness of $\mathbb{K}$ can be illustrated with a complete formalization of an ERC20
token contract~\cite{ERC20K}.
Despite different choices of a verification technique, the aforementioned approaches
are able to capture the dynamic aspects of smart contract execution, such as
the memory state and gas consumption.

Such verification systems are usually built upon formal semantics of smart
contract languages, e.g., EVM bytecode~\cite{Hirai2017,Amani2018,Duo2020}, Solidity~\cite{Yang2019a},
Michelson~\cite{Bernardo2019}. To make contracts amenable to formal reasoning,
developers equip smart contract languages with a logical foundation,
such as $\lambda$-calculus, and formal semantics~\cite{Sergey2019,Chapman2019}.

At the same time, smart contracts that are written or translated to Java can be verified
using a dynamic logic called JavaDL---a program logic for Java. A Java program annotated
with JML specifications is automatically translated to JavaDL by KeY---a framework
for deductive verification.
While Hyperledger Fabric contracts written in Java can be verified directly~\cite{Beckert2018,Beckert2019},
Solidity smart contracts are preliminarily translated to Java~\cite{Ahrendt2019a}.
Dynamic logic can be seen as an extension of Hoare logic and is also used
to define an executable specification for smart contracts~\cite{Sato2018}.
The concepts related to dynamic logic are discussed in more detail in~\cref{sssec:contract-specification}
and~\cref{sssec:program-specification}.

Some researchers consider logic-based languages to be particularly useful for the
implementation and verification of \emph{legal smart contracts}~\cite{Governatori2018}.
A legal smart contract is a digitized representation of a legal agreement, which typically
includes deontic modality statements describing obligation, permission, and prohibition of the parties.
To capture contractual aspects, the authors of~\cite{Governatori2018}
formalize a smart contract in a logic-based language called \emph{Formal Contract Logic} (FCL).
Azzopardi et al.~\cite{Azzopardi2018} propose a formal deontic logic
with small-step operational semantics for Solidity smart contracts~\cite{Azzopardi2018}.
Specifications defined in deontic and defeasible logics are discussed
in~\cref{sssec:contract-specification}.

\subsection{Summary}\label{ssec:summary-forms}
The analyzed work illustrates different aspects of smart contracts that researchers
focus on, together with the appropriate formalisms.
As such, the concurrent nature of smart contract execution is captured by contract-level
models based on process algebra and Petri net, while game-theoretical structures are suitable
for modeling a multi-agent environment.
Contract-level models, and state-transition systems, in particular, provide an intuitive
way to outline high-level behavior of a smart contract, which includes possible interactions with other
contracts, users, or blockchain.
Still, the abstraction over the execution logic of a smart contract introduces a gap between
an original contract and its representation in a chosen formalism.
On the other hand, program-level representations
give intuition about the lower-level details of the execution process.
At the same time, while bytecode-level modeling provides wide capabilities in terms of low-level
reasoning, e.g., about resource consumption,
it also complicates specification of semantic properties related to smart contract
behavior~\cite{Antonino2020}.

%% file: specifications.tex
\section{Survey on Smart Contract Specifications}\label{sec:specifications}
In this section, we review the literature on smart contract specifications.
Requirements on smart contract applications are usually specified as \emph{properties}---statements
written in some formal languages about program behaviors.
In \cref{ssec:form-specifications}, we survey the formalisms adopted in formal specification,
which can be categorized into the contract- and program-level specifications.
In \cref{ssec:domain-properties}, we review various types of properties (e.g., security and
functional correctness) from different application domains, observed in the published work under
study.

\subsection{Formal Specifications for Smart Contracts}\label{ssec:form-specifications}
Similar to the modeling formalisms discussed in~\cref{sec:modeling}, we make a distinction between
\emph{contract-level} and \emph{program-level} formal specifications of smart contracts.
This is in correspondence with the traditional classification of \emph{model-oriented} and
\emph{property-oriented} specifications, rooted from \emph{temporal logic} and \emph{Hoare logic},
respectively.
We discuss contract-level specifications about high-level contract properties
in~\cref{sssec:contract-specification} and program-level specifications governing lower-level
program implementations in~\cref{sssec:program-specification}.

\subsubsection{Contract-Level Specification}\label{sssec:contract-specification}
Contract-level specifications of smart contracts are expressed in various
types of logic.
The most prevalent category is a family of temporal logics, which includes linear and branching
temporal logics and their dialects.
Other logics include deontic and defeasible logics, which help to define the rights
and obligations of smart contracts' parties in accordance with a legal contract.
As specifications in temporal~\cite{Mavridou2019,Suvorov2019}, dynamic~\cite{Sato2018},
and deontic logics~\cite{Frantz2016,Boogaard2018} capture high-level behavioral characteristics of
a system, they are used to synthesize correct-by-design smart contracts.
Propositional and predicate logics usually convey factual statements about the contract's state over the traces.

\paragraph{Temporal Logics}
Temporal logics express properties of a smart contract over time.
The execution logic of smart contracts often encloses time constraints, be it
legal contractual clauses~\cite{Clack2018} or ending time of an auction or voting contract~\cite{Frantz2016}.
As noted in~\cite{Nehai2018}, in relation to smart contracts, the word
``temporal'' may refer to logical time considering only the order of events.
Defined over a sequence of states, temporal formulae constitute a suitable language to
describe the behavior of a state-transition system---a common formal representation
of a smart contract described in~\cref{sssec:transition-systems}, which is then
verified by a model checker.

The two key temporal properties are \emph{safety} and \emph{liveness}.
Safety properties correspond to a statement: ``Nothing bad ever happens'',
and are often used to express invariance.
In addition to general software properties such as deadlock-freedom,
common safety properties of smart contracts define permissible values of state variables,
establish invariants over a smart contract balance (C.f.~\cref{sssec:properties-financial}),
specify access control for functions, or define a possible order of events.

Liveness properties state that ``something good eventually happens''.
They characterize the possibility of a program to progress, which, for a smart contract,
is often reflected in its ability to give away cryptocurrency.
The primary liveness property is \emph{liquidity}~\cite{Bartoletti2019}--a security property discussed
in detail in~\cref{sssec:properties-security}.
Other liveness properties specify the ability of a smart contract or its users to eventually
reach a particular desired state~\cite{Madl2019,Nehai2018}, even considering different user
strategies~\cite{Meyden2019}.
Unlike a safety property, however, there exists no finite counterexample for a liveness
property. To make them amenable to symbolic analysis, some authors~\cite{Nikolic2018,Shishkin2018}
define under-approximations of liveness properties that are checked on traces of
a fixed length.

\emph{Linear Temporal Logic} (LTL) is a fragment of first-order logic (FOL) which
is used to define properties over a linear sequence of successive states of a
program. LTL properties describe safety and liveness of transition-system
models of smart contracts that are verified by a model checker~\cite{Bai2018,Molina-Jimenez2018,Atzei2019,Bartoletti2019,Alqahtani2020}.
Temporal safety properties of the \contract{Blind Auction} smart contract in LTL proposed by Suvorov
et al.~\cite{Suvorov2019}
are depicted in~\cref{fig:auction_properties}. The properties are also relevant
for the \contract{Simple Auction} smart contract, which is formally modeled in~\Cref{sec:modeling}
(\cref{fig:auction_code,fig:auction_fsm}).
\emph{Past-time LTL} (ptLTL) has a set of temporal operators that reference
the past states of a trace and allow succinct formulae for safety properties~\cite{Permenev2020,BogdanichEspina2019}.
For example, two ptLTL-based specification languages for Ethereum~\cite{Permenev2020,BogdanichEspina2019}
support an operator \emph{previously}, which refers to the value of a contract variable
before the transaction execution. The properties defined in these languages are then translated to source-code
assertions.
To account for the nondeterministic execution environment of a smart contract,
Abdellatif et al.~\cite{Abdellatif2018b} specify properties in \emph{Probabilistic Bounded LTL (PB-LTL)}.
Probabilistic properties in~\cite{Abdellatif2018b} estimate the probability of a successful attack on a
name registration contract on each step of a transaction lifecycle.

\begin{figure}[tb]
    \footnotesize
    \begin{minipage}[c]{0.45\linewidth}
        (i) \texttt{bid} cannot happen after \texttt{close}:\\
        \centerline{$\mathbf{G}\;(\mathtt{close}\rightarrow\neg\mathbf{F}\;\mathtt{bid})$}\\
        (ii) \texttt{withdraw} cannot happen before \texttt{finish}:\\
        \centerline{$\mathtt{finish}\;\mathbf{R}\;\neg\mathtt{withdraw}$}
        \end{minipage}
        \begin{minipage}[c]{0.45\linewidth}
        (i) \texttt{bid} cannot happen after \texttt{close}:\\
        \centerline{$\mathbf{AG}\;(\mathtt{close}\rightarrow\mathbf{AG}\;\neg\mathtt{bid})$}
        (ii) \texttt{withdraw} can happen only after \texttt{finish}:\\
        \centerline{$\mathbf{A}\;[\neg\mathtt{withdraw}\;\mathbf{W}\;\mathtt{finish}]$}
        \end{minipage}
    \caption{Properties of an {\contract{auction}} smart contract in LTL from~\cite{Suvorov2019} (left)
        and CTL from~\cite{Mavridou2019} (right).}\label{fig:auction_properties}
    \end{figure}

\emph{Computation Tree Logic} (CTL) is a branching-time logic used to describe properties of a computation tree.
In contrast to LTL formulae, whose semantics model considers one path (i.e., a sequence of
states) without branching, CTL operators quantify over a tree of all possible future (branching)
paths starting from the given state.
Still, CTL is also used to formalize the specification of an FSM
representation of a smart contract for model checking~\cite{Nehai2018,Mavridou2019,Madl2019}.
\Cref{fig:auction_properties} also illustrates a CTL formalization of temporal properties for
an \contract{auction} smart contract proposed by Mavridou et al.~\cite{Mavridou2019}.
As in LTL, various dialects of CTL are used to formulate specific
requirements of smart contract models. An extension of CTL for Colored Petri Nets (CPN),
\emph{ASK-CTL}, is used to describe correctness of CPN models of smart contracts~\cite{Liu2019a,Duo2020}.
To reason about the quantitative aspect of time in a timed commitment protocol,
Andrychowicz et al.~\cite{Andrychowicz2014} use a \emph{Timed CTL} (TCTL) language of
the \tool{UPPAAL} model checker to define properties of the timed automata models.
\emph{Probabilistic CTL} (PCTL) properties of a Markov Decision Process %
model of a selling contract measure the probability of one of the players
to lose a certain percentage of his wealth after a given amount of time~\cite{Bigi2015}.
\emph{Alternating Temporal} (ATL) and \emph{Strategic Logics} (SL)
are multi-agent extensions of CTL that allow quantification over agents'
strategies. They describe the behavior of a concurrent game model of an atomic swap~\cite{Meyden2019}.

\paragraph{Dynamic Logic}
\emph{Dynamic logic} allows formulating properties it terms of \emph{actions} taking the termination of a program into consideration.
The authors of \cite{Sato2018} propose a DSL for smart contract implementation
based on properties written in \emph{Linear Dynamic Logic on finite traces ($LDL_f$)}.
Owing to its better expressiveness compared with LTL, LDL$_f$ is used directly to model a
smart contract, instead of encoding the state-transition model in a separate modeling language,
such as Promela.
Other applications of dynamic logic are related to deductive verification of smart contracts
in the \tool{KeY} framework. However, the properties are formulated
in a smart contract as JML annotations~\cite{Ahrendt2019a,Beckert2018,Beckert2019}
and are then translated to Java Dynamic Logic (JavaDL) (C.f.~\cref{sssec:program-specification}).

\paragraph{Deontic and Defeasible Logics}
\emph{Deontic logic} is a division of modal logic operating over the concepts of obligation, permission, and prohibition.
As legal prose relies on deontic modality~\cite{Ladleif2019}, deontic-logic specification
of a legal smart contract can bridge the gap between its legal semantics and implementation~\cite{Clack2018}.
However, deontic modality requires special treatment from the specification
perspective, as permission has no direct correspondence in many specification
languages~\cite{Azzopardi2018,Suvorov2019}.
To enable compliance checking between a legal contract and a Solidity smart contract,
Azzopardi et al.~\cite{Azzopardi2018} proposed an operational semantics for contracts
written in deontic logic. Deontic statements also constitute one of the components of
\emph{ADICO institutional grammar}~\cite{Crawford1995}, which captures institutional rules, conventions, and norms.
Given a set of ADICO rules, a framework proposed in~\cite{Frantz2016}, can generate a
skeleton of a Solidity smart contract.

\emph{Defeasible logic} is a non-monotonic logic which allows for so-called
defeasible propositions, in other words, exceptions to some of the
rules. Such specifications contain a statement that regulates superiority between the
clauses of a contract, which helps to handle possible disputes and breaches.
The authors of~\cite{Governatori2018} encode a smart legal contract that regulates a product license evaluation
into a set of logical rules in \emph{Formal Contract Logic} (FCL)---a formal contractual language
based on defeasible logic and deontic logic of violations.
The authors note that such logic-based implementation can be seen as an executable specification
and, hence, facilitates monitoring.

\paragraph{Other Logics}
We found several approaches that use other types of logic to describe contract-level
requirements for smart contracts. For example, Li et al.~\cite{Li2019bnb} analyze traces of a process-algebraic
contract model based on properties written in a fragment of (temporal) FOL.
Temporal properties over traces of a \emph{Scilla} smart contract
are formalized using Coq higher-order logic predicates and a proposed
temporal connective~\cite{Sergey2018}.

Unal et al.~\cite{Unal2019} use a \emph{formal role-based access control (FPM-RBAC)} security
model to specify the behavior of a crypto-wallet smart contract in a 5G network environment.
The specification consists of role-based access control statements, generic constraints expressed
in predicate logic, and \emph{ambient logic} constraints for location- and time-based access policy.

\subsubsection{Program-Level Specification}\label{sssec:program-specification}
A dual type of smart contract specification is defined at the program level.
In this section, we concentrate on Hoare-style specifications, which are often
used together with program logics (C.f.~\cref{sssec:program-logic}) or are instrumented in the actual smart contract source code. Then, we review
specifications that rely on predefined patterns of instructions and events
observed during smart contract execution.

\paragraph{Hoare-Style Properties}
A \emph{Hoare-style} property of a program $c$ is captured by
a triple $\models \{P\} c \{Q\}$, where $P$ and $Q$ are predicates on the state
of a program and its execution environment~\cite{Amani2018}.
$P$ and $Q$ describe the state before and after the execution of $c$
(if it terminates), respectively, and are referred to as a \emph{precondition} and a \emph{postcondition}.
An \emph{invariant} here is a predicate preserved by the execution of program $c$.
Generally, Hoare triples express \emph{partial correctness}, while
\emph{total correctness} requires proof of program termination.
When applied to smart contracts, partial and total correctness are often considered equivalent
due to the gas mechanism of Ethereum: a contract is guaranteed to terminate either successfully or due to an out-of-gas
exception~\cite{Amani2018,Genet2020}.

A Hoare-style specification of a smart contract consists of preconditions, postconditions,
and invariants defined for its functions, loops, and the contract in general.
Predicates in these statements include storage and memory variables,
environmental variables capturing block or transaction information, events and operations emitted during
the execution. Hoare-style properties, therefore, can capture both vulnerable conditions and semantic
correctness of smart contract functions.
For example, access control in smart contracts is often realized via a precondition check
on the identity of a transaction sender.
As another example, postconditions are used to check if the constructor has initialized
the contract state properly~\cite{Wang2019Azure,BogdanichEspina2019}.
Hoare-style properties can be specified using program logics (C.f.~\cref{sssec:program-logic})
and are often supported by verification languages, such as Why3~\cite{Nehai2019,Zhang2020Why3}.
Alternatively, Hoare-style specifications can be defined as smart contract assertions,
that are analyzed by verification tools
statically~\cite{Albert2019,Permenev2020,Hajdu2019,Wang2019Azure,Kalra2018,Akca2019,So2019,Alt2018}
or in runtime~\cite{Li2020,Stegeman2018,BogdanichEspina2019}.

Many program logics listed in~\cref{sssec:program-logic} enable Hoare-style reasoning
about smart contracts in theorem provers~\cite{Li2019,Bernardo2019,Amani2018,Yang2019a}.
The work in~\cite{Amani2018} demonstrates how this approach is used to
describe correctness of an escrow smart contract.
To consider all possible invocations, the authors also verify that any input that is
not covered by a specification leads to a transaction being reverted.

Nehai et al.~\cite{Nehai2019} define a Hoare-style specification of smart contract functions
in the polymorphic FOL of \tool{Why3}---an instrument for deductive verification.
The specification contains requirements for both semantic correctness and lower-level
operational details, such as gas consumption and the ordering of array elements.
Other techniques based on deductive verification~\cite{Ahrendt2019a,Beckert2018,Beckert2019}
use the \tool{KeY} framework. In this case, the input consists of a Java smart
contract annotated with \emph{Java Modeling Language} (JML) properties expressing pre-/postconditions,
invariants, and modifiable variables of contract functions.
The authors of~\cite{Beckert2019}, however, note the inability of their
approach to handle complex properties, including temporal, even if
they are expressible in JML. JML specification is also supported by
\tool{Raziel}~\cite{Sanchez2018Raziel}---a framework for the implementation of
\emph{proof-carrying} smart contracts. \tool{Raziel} supports JML for functional
correctness annotations and \emph{Obliv-Java}---for privacy variable annotations.
A proof-carrying code proposal for safe smart contract upgrade by Dickerson et al.~\cite{Dickerson2019}
also relies on a Hoare-triple specification.

Assertions that correspond to Hoare-style properties can be added to a smart contract
by means of Solidity. The language includes \texttt{require} and \texttt{assert} statements
that can express a precondition or an invariant and postcondition, respectively.
\texttt{require} statements are widely used to validate
the input received from a user---the absence of such checks is
known to be a source of serious issues~\cite{Groce2019,Cecchetti2020}.
To advance further program-based specification, several tools provide
custom specification languages~\cite{Li2018,Permenev2020,Hajdu2019,Wang2019Azure,Li2020,Steffen2019,Stegeman2018,BogdanichEspina2019}.
They support quantifiers~\cite{Li2020}, the implication
operator~\cite{Permenev2020,BogdanichEspina2019}, and auxiliary functions,
such as \emph{sum}, which refers to a sum of values stored in a mapping~\cite{Permenev2020,Hajdu2019,Li2020}.
The latter is useful in specification of financial smart contract invariants,
as shown in~\cref{sssec:properties-financial}.
To check semantic correctness of a smart contract,
\tool{VeriSol}~\cite{Wang2019Azure} translates a state-machine policy
of a Solidity smart contract into Hoare-style source code assertions.
It can, therefore, identify an incorrect state transition and an incorrect initial state in a smart contract.
Apart from user-defined specifications, several tools detect smart contract vulnerabilities
by adding assertions to a smart contract automatically~\cite{Akca2019,So2019}.
Overall, instrumentation of a smart contract code with the corresponding checks introduces a transparent
and consistent approach to smart contract specification.
However, while intended to be side-effect free expressions, source-code assertions generate
executable code and, therefore, gas consumption.

\begin{figure}[t]
\begin{minipage}{.37\textwidth}
  \subfigure[Unchecked Send]{
    \includegraphics[width=.85\textwidth]{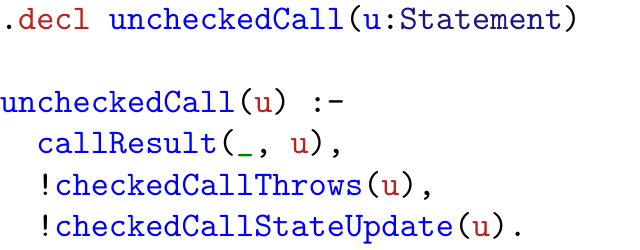}\label{fig:send_pattern}}
  \subfigure[Accessible Selfdestruct]{
    \includegraphics[width=.85\textwidth]{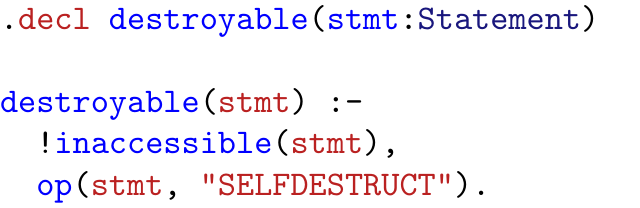}\label{fig:destroyable_pattern}}
  \caption{Patterns for ``Unchecked Send'' and ``Accessible Selfdestruct'' vulnerabilities detected by
  \tool{Vandal} in~\cite{Brent2018}.}
  \label{fig:vandal_patterns}
\end{minipage}
\quad
\begin{minipage}{.52\textwidth}
  \includegraphics[width=\textwidth]{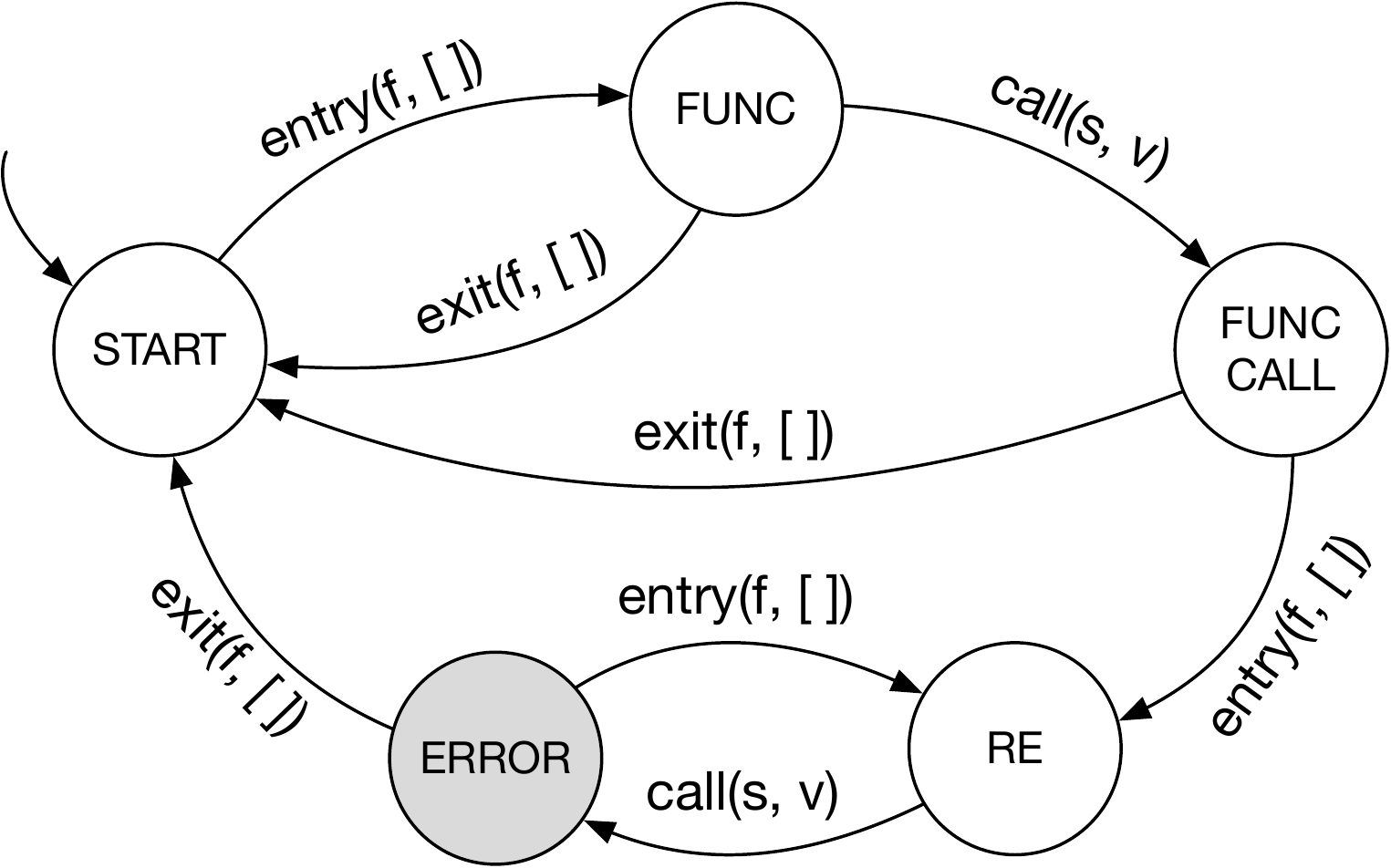}
  \caption{An automaton for identification of reentrancy during execution.
  Reproduced from~\cite{Liu2018Reguard} with permission.}
  \label{fig:reentrancy_pattern}
\end{minipage}
\end{figure}

\paragraph{Program Path-Level Patterns}
Symbolic execution represents each program trace as a propositional formula
over symbolic inputs to a program required to follow this execution path.
Based on a contract CFG, this approach allows recognizing traces that contain
problematic sequences of program states and instructions. In smart contract
verification, the existing applications indicate if the contract is
vulnerable~\cite{Luu2016,Torres2018,Zhou2018,Chang2018,Nikolic2018,Mueller2018,He2020,Quan2019},
gas-inefficient~\cite{Chen2017}, or a honeypot~\cite{Torres2019}.
Some authors encode patterns over instructions and flow- and data-dependency
in verification languages, such as \emph{Datalog}, to certify contract security~\cite{Tsankov2018,Brent2018} including the
absence of gas-related vulnerabilities~\cite{Grech2018}.
\Cref{fig:vandal_patterns} presents patterns that correspond to \emph{Unchecked Send}
and \emph{Accessible Selfdestruct}~\cite{Brent2018} vulnerabilities.
The topmost pattern detects a vulnerability if the result of an external call is not used to update
a blockchain state or throw an exception, while the bottom one detects a \texttt{selfdestruct}
operation that can be called by an arbitrary account (C.f.~\cref{sssec:properties-security}).
The code snippets illustrating both vulnerabilities are shown in~\cref{fig:vulnerable-patterns-koet}.

While the pattern-based approaches discussed above operate on an EVM compilation of a smart
contract, Perez and Livshits~\cite{Perez2019} use \emph{Datalog} to formulate vulnerable
patterns over execution traces retrieved from the blockchain.
During fuzzing, \tool{ReGuard}~\cite{Liu2018Reguard} compares an execution
trace of a smart contract to an automaton that captures reentrancy.
The automaton is shown in~\cref{fig:reentrancy_pattern}, where \texttt{entry} and \texttt{exit}
correspond to an entry and exit of a function, and \texttt{CALL} is the corresponding EVM
instruction.
Another common dynamic vulnerability detection technique identifies predefined patterns of
instructions in runtime~\cite{Gao2019,Ma2019,Rodler2018}.

Another type of path-level specification is based on events.
Since the ERC20 standard specifies the events emitted by contract functions, several techniques
perform compliance checking considering events too~\cite{Hajdu2019,Chen2019TokenScope}.
Events emitted by a smart contract during its execution are analyzed
for compliance with business rules by Fournier et al.~\cite{Fournier2019}
and Molina-Jimenez et al.~\cite{Molina-Jimenez2018}. The former specification
ensures acceptable transportation conditions in a supply chain use case,
while the latter encodes clauses of a selling contract in terms of parties' responsibilities.
Shishkin~\cite{Shishkin2018} formalizes functional requirements of an Ethereum
contract as predicates over traces of events.

In addition, some publications adopt a different notion of events.
For example, \tool{EthRacer}~\cite{Kolluri2019} identifies event-ordering bugs
in smart contracts by comparing results of different permutations of function
invocations. Therefore, in this work, an event refers to a function call.
The authors of~\cite{Grossman2017} verify serializability of smart contract execution
by analyzing its traces of events, or transitions between contract states.
The corresponding property of being \emph{effectively callback free} (ECF)
is satisfied if for all executions containing a callback there exists
an equivalent execution without callback that can start in the same state and
end in the same final state.
A vulnerable pattern of events corresponding to actions performed by a CSP model
(C.f.~\cref{sssec:process-algebras}) of a smart contract helps to detect an attack
in its execution~\cite{Qu2018}.

\subsection{Properties Classification by Domains}\label{ssec:domain-properties}
Using the listed formalisms, specifications describe smart contracts' functional correctness from
different perspectives. The considered aspects include the absence of vulnerabilities,
respect for privacy requirements, reasonable resource consumption, conformance to
business-level rules, fairness to users.
This section begins with an overview of existing classifications of domain-specific
smart contract properties.
Then, based on the property domains, we propose our categorization of
smart contract properties that we have frequently seen in the research literature.

\paragraph{Existing Smart Contract Property Classification}
Several taxonomies were proposed for smart contract
vulnerabilities~\cite{Atzei2017,Yamashita2019,Groce2019,Sayeed2020,Chen2019a},
with the first and, arguably, the most influential published in 2017~\cite{Atzei2017}.
\Cref{sssec:properties-security} overviews security properties that
serve to protect a smart contract from types of vulnerabilities included in
these classifications.

Two other publications describe common classes of domain-specific properties
for smart contracts~\cite{Permenev2020,Bernardi2020}. These articles
adopt dual forms of property specification: Hoare triples and temporal properties,
respectively. Nevertheless, their categorizations share several
types, such as \emph{user-based access control} over contract functions and \emph{invariants over
aggregated values of collections}. In addition, Permenev et al.~\cite{Permenev2020}
define transition-system types of properties for smart contracts and extend
invariants to a multi-contract setting.
Some of these properties also resemble software design patterns~\cite{Wohrer2018},
which capture the recurrent problems in smart contract development, e.g.,
\emph{State Machine}, \emph{Access Restriction}, \emph{Ownership}~\cite{Wohrer2018}.
Bernardi et al.~\cite{Bernardi2020}, on the other hand,
consider two types of properties related to a cryptocurrency exchange.
We review some of these properties in~\cref{sssec:properties-financial}.

Our domain-based analysis of properties is also shaped by a catalog of
smart contract application domains proposed by Bartoletti et al.~\cite{Bartoletti2017}.
They divide smart contracts into \emph{Financial, Notary, Game, Wallet, and Library}
types. In~\cref{ssec:domain-properties}, we describe our findings on the properties
that are common for some of these types of smart contracts.

\subsubsection{Security}\label{sssec:properties-security}
Detection of security flaws in smart contracts received a lot of attention.
Important security properties studied by the research community include \emph{liquidity} (``a
non-zero contract balance is always eventually transferred to some
participants~\cite{Nikolic2018,Bartoletti2019}''),
\emph{atomicity} (``if one part of the transaction fails, then the entire transaction fails and the
state is left unchanged''~\cite{Luu2016}), \emph{single-entrancy} (``the contract cannot perform
any more calls once it has been reentered''~\cite{Schneidewind2020}), \emph{independence of the
mutable state and miner-controlled parameters}~\cite{Grishchenko2018,Wang2019payment}.
Security guarantees of a smart contract also include proper \emph{access
control}~\cite{Tsankov2018,Nikolic2018} for safety-critical operations,
\emph{arithmetic correctness}~\cite{So2019,Feng2019a,Kalra2018}, and
\emph{reasonable resource consumption}~\cite{Bhargavan2016,Chen2017,Grech2018}.
The similarity of a smart contract to a concurrent object~\cite{Sergey2017} suggests that some
security guarantees can be derived from checking for \emph{linearizability}~\cite{Kolluri2019} and
\emph{serializability}\cite{Grossman2017} of its executions.

\begin{figure}[t]
  \begin{minipage}{.48\textwidth}
      \includegraphics[width=\textwidth]{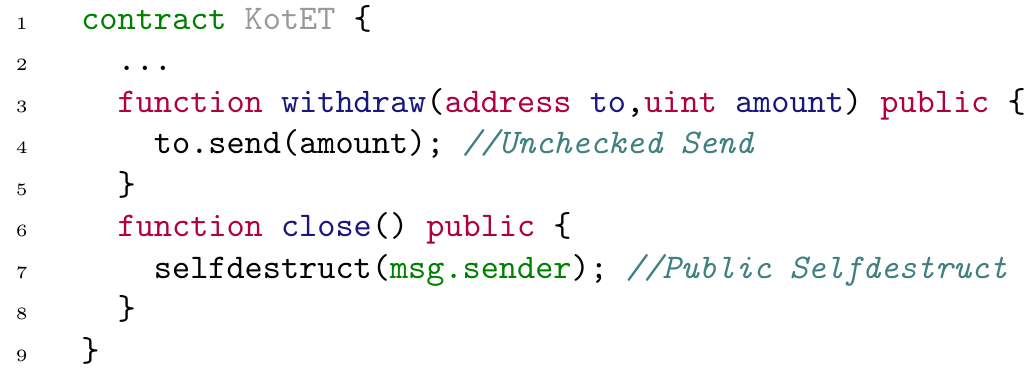}
      \caption{``Unchecked Send'' and ``Accessible Selfdestruct'' vulnerabilities~\cite{Wang2019VUL}.}
      \label{fig:vulnerable-patterns-koet}
  \end{minipage}
  \quad
  \begin{minipage}{.48\textwidth}
    \includegraphics[width=\textwidth]{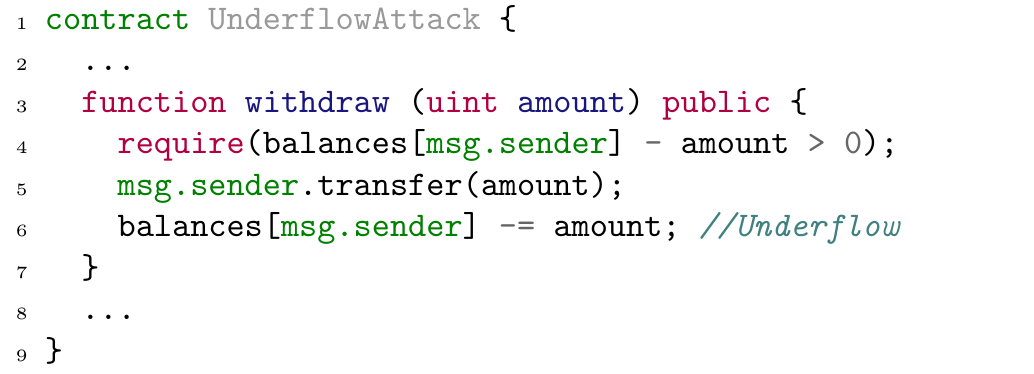}
    \caption{``Integer Underflow'' vulnerability~\cite{Underflow}.}
    \label{fig:vulnerable-patterns-underflow}
  \end{minipage}
\end{figure}

The violation of these properties leads to many well-known smart contract vulnerabilities.
For example, contracts violating liquidity are called ``\emph{greedy}'', implying that their funds
cannot be released under any conditions~\cite{Nikolic2018}.
A smart contract violates atomicity if it does not check the return value of a (possibly failed)
external call operation, which corresponds to the \emph{unchecked call}
vulnerability~\cite{Grishchenko2018,Perez2019}.
The execution logic of a smart contract that is not independent of environmental variables,
e.g., \texttt{block.timestamp}, is prone to \emph{dependence manipulation}~\cite{Wang2019payment}
by a miner, including the \emph{timestamp dependency} vulnerability~\cite{Luu2016}.
If the business logic of a smart contract depends on its mutable state parameters, such as balance
and storage, then it has the \emph{transaction-ordering dependence} (TOD) problem.
TOD belongs to a class of \emph{event-ordering}~(EO) bugs, which arise from the nondeterministic
execution of blockchain transactions~\cite{Kolluri2019} and are manifested by different outputs
produced by different orderings of the same set of function invocations~\cite{Kolluri2019,Wang2019payment}.
The authors of~\cite{Kolluri2019} use the notion of linearizability to identify EO bugs, including
mishandled asynchronous callbacks from an off-chain oracle.
The violation of single-entrancy implies that a smart contract is \emph{reentrant}:
provided with enough gas, an external callee can repeatedly call back into a smart contract within
a single transaction.
In execution traces, reentrancy is also identified as a violation of a serializability-like
property~\cite{Grossman2017}, i.e., if the result of an execution with external callbacks is not
equivalent to that of an execution without.
Missing permission checks for the execution of a transfer or a \texttt{selfdestruct} operation
make a smart contract \emph{prodigal} and \emph{suicidal}~(\cref{fig:vulnerable-patterns-koet}), respectively~\cite{Nikolic2018}.
The identification of a transaction sender through the \texttt{tx.origin} variable is also considered
vulnerable~\cite{Brent2018}. Integer arithmetics of EVM and absence of automatic checks for arithmetical
correctness make Ethereum contracts prone to \emph{integer overflow and underflow}~\cite{So2019,Feng2019a}.
\Cref{fig:vulnerable-patterns-underflow} presents smart contract code containing the \emph{integer underflow}
vulnerability.
Furthermore, the progress of a smart contract can be compromised by gas-exhaustive code
patterns~\cite{Chen2017,Grech2018}.

Many symbolic execution tools detect vulnerabilities by scanning
for vulnerable patterns in path conditions and traces derived from a contract CFG.
For example, to identify \emph{timestamp dependency}, \tool{Oyente}~\cite{Luu2016} checks if a path
condition of a trace includes the \texttt{timestamp} variable.
To detect \emph{suicidal} smart contracts, \tool{sCompile}~\cite{Chang2018} checks whether, for the
paths that contain \texttt{selfdestruct}, the path conditions constrain on the timestamp, block
number, or the address of the contract owner.
\tool{Maian}~\cite{Nikolic2018} flags traces which transfer funds or control to an address
not appearing in a contract state, as \emph{prodigal}.
Scanning for vulnerable instruction patterns can also be done in runtime, for example,
to identify that a permission check relies on \texttt{tx.origin}~\cite{Chen2020SODA}.

He et al.~\cite{He2019} consider prodigal and suicidal contracts to be the most
challenging to detect, as it requires an ordered set of transactions.
Indeed, if one sets the \texttt{owner} variable of a smart contract equal to his address,
he can bypass the aforementioned checks in the following transactions. The authors of~\cite{Brent2020,Cecchetti2020}
call such vulnerabilities \emph{composite} and detect them using information-flow patterns.
Other static analysis tools also detect patterns over the data- and control-flow of a smart contract, capturing
the necessary context-sensitive information.
\Cref{fig:send_pattern} demonstrates a pattern for an \emph{unchecked call} vulnerability
detection used by \tool{Vandal}~\cite{Brent2018}.
The pattern states that the return value of the \texttt{call}
operation should be used either to control the execution of the \texttt{throw} operation or to perform
storage writes. To prevent \emph{overconsumption of gas}, static analyses
infer the upper bound on gas consumed by a function call~\cite{Bhargavan2016,Nehai2019,Albert2019Gasol,Marescotti2018}
or detect vulnerable patterns, such as unreachable code~\cite{Chen2017}, trivial conditions~\cite{Alt2018}
or an unbounded number of elements~\cite{Grech2018} in loops.
Race conditions, such as TOD, are identified by static analyses as read-write hazards between the executions of
two~\cite{Tsankov2018,Kalra2018,Grishchenko2018} or several~\cite{Wang2019payment} functions.
An EO bug caused by an asynchronous callback could as well be prevented by a temporal property proposed in~\cite{Sergey2018}:
it demands the state of a contract to remain unchanged until the response from an oracle is received.

As a liveness property, \emph{liquidity} can be intuitively formulated in temporal logics~\cite{Bartoletti2019}.
Bartoletti et al.~\cite{Bartoletti2019} additionally extend its definitions
to user strategies: the default, \emph{strategyless multiparty liquidity}, assumes the existence
of a cooperative strategy to unfreeze funds.
Liquidity can also be compromised by DoS attacks,
for example, \contract{Parity Multisig Wallet}~\cite{Parity2017Frozen} lost
its ability to pay users as it delegated transfers to a smart contract that was destroyed.
Nelaturu et al.~\cite{Nelaturu2020} identify the \contract{Parity} issue as a violation of deadlock-freedom.

Since arithmetic issues are often caused by non-validated function arguments,
their identification is entrusted to pre- and postconditions~\cite{Hirai2017,Nehai2019,Park2018,Kasampalis2019}
and assertions~\cite{So2019,Feng2019a}.
The (formally certified~\cite{Schneidewind2020,Zhang2020Why3}) \texttt{SafeMath}~\cite{SafeMath}
Solidity library implements similar checks via the \texttt{require} statements.
Source-code assertions are also a popular way to implement role-based \emph{access control} in
smart contracts.
For example, it is done via equality checks between the values of \texttt{msg.sender}
(the usage of \texttt{tx.origin} is considered vulnerable~\cite{Chen2020SODA,Brent2018})
and the corresponding contract variable, e.g., \texttt{owner}~\cite{Permenev2020,Wang2019Azure,Weiss2019,Hajdu2019}.
Functions with restricted access usually change the value of a safety-critical variable, such as \texttt{owner},
or manipulate funds, e.g., transfer or mint tokens.

\emph{Reentrancy} is one of the most studied smart contract vulnerabilities.
Both static and dynamic data- and control-flow analyses
detect the violation of \emph{``no writes after call''} and similar
patterns~\cite{Tsankov2018,Feng2019a,Li2018,Torres2019aegis,Rodler2018,Lu2019}
(yet, this pattern is neither sound nor complete approximation of reentrancy~\cite{Schneidewind2020}
and is considered a standalone property in~\cite{Tsankov2018}).
A similar check is specified in terms of contract operations in CTL~\cite{Mavridou2019}.
While \emph{single-entrancy} is verified statically using Horn clauses in~\cite{Schneidewind2020},
a similar property is also defined in terms of instructions in the execution
trace~\cite{Liu2018Reguard} (\cref{fig:reentrancy_pattern}).

Finally, the inconsistency caused by reentrancy as well as other issues, such as flawed
arithmetics~\cite{Li2020}, can be detected through invariants.
Such invariants are proposed for financial~\cite{Li2019bnb,Permenev2020,Shishkin2018,Hirai2018}
(C.f.~\cref{sssec:properties-financial})
and voting contracts~\cite{Nielsen2019} (C.f.~\cref{sssec:properties-social}).

\subsubsection{Privacy}\label{sssec:properties-privacy}
Smart contracts that implement auctions, games, and lotteries select a winner among the users who
submitted their bids or moves.
Blockchain transactions are publicly accessible, so a malicious user can decide on his move based
on the actions of other contract participants.
In a name registration smart contract, one can intercept the name chosen by another user from
an unconfirmed transaction and register his address under this name first~\cite{Abdellatif2018b}.

To prevent exposure of user inputs in a smart contract by design, it is recommended to implement
cryptographic protocols, such as \emph{commitment}~\cite{Delmolino2016} and
\emph{timed commitment}~\cite{Bartoletti2019a} schemes.
The protocol requires each participant to commit to a secret that corresponds to his
input and should be revealed in time---otherwise, the participant has to pay
compensation.
In general, the correctness of a commitment smart contract is described in terms of the knowledge of
each others' secrets that participants did or did not obtain as well as the corresponding changes
in their balance. One example of such a property is proposed by the authors
of~\cite{Andrychowicz2014}: \emph{``if an honest Bob did not learn Alice's secret then he gained
Alice's deposit as a result of the execution''}.
Similar aspects are considered in specifications of \emph{atomic-swap} contracts,
which allow two parties to exchange their assets \emph{atomically}, i.e., the swap either
succeeds or fails for both parties~\cite{Meyden2019,Hirai2018}.
Alternatively, to prevent contract participants from cheating, Sergey et al.~\cite{Sergey2018}
suggested formulating a property as a knowledge argument over the prefix of observed execution
history.

The transparency of the transactions and ledger data also makes smart contracts less applicable
to privacy-sensitive domains, such as healthcare and voting.
Potential solutions for privacy issues in smart contracts include the application of trusted
management~\cite{Kosba2016} or hardware~\cite{Zhang2016,Cheng2019}.
On the whole, privacy-preserving smart contract is an independent topic of intense study, not
covered by this survey.
Among these approaches, we distinguish two that enforce privacy guarantees by allowing the
corresponding annotations in smart contract code.
To prevent data leaks, a smart contract language \tool{zkay}~\cite{Steffen2019} supports variable
annotations, which specify who is eligible to read the value of a variable.
Similar privacy annotations are supported by the \tool{Raziel}~\cite{Sanchez2018Raziel} framework.
As a demonstration, a smart contract is implemented in the Obliv-Java programming language with
both privacy and correctness requirement annotations.
Both tools utilize some zero-knowledge proof protocols to certify that privacy is maintained in a
smart contract before it is executed.

\subsubsection{Finance}\label{sssec:properties-financial}
One of the key advantages of smart contracts is their ability to manipulate digital assets, known
as cryptocurrency.
Smart contracts automate the execution of financial applications that involve storage and transfer
of funds, between users and contracts.
Common examples include wallets, banks, escrows, cryptocurrency exchanges, and crowdfunding
campaigns.

\emph{Initial Coin Offering} (ICO) is a method of raising funds by selling
\emph{tokens}---programmable assets managed by smart contracts on blockchain
platforms~\cite{Chen2020Token}.
Analysis performed by Chen et al.~\cite{Chen2020Token} demonstrates that 80\% of ICOs
are Ethereum-based smart contracts implemented according to the ERC20 standard~\cite{ERC20}.
The standard describes an interface of a token contract, i.e., its functions and
events. Arguably, even this semi-formal specification gave rise to the application
of formal techniques to token
contracts~\cite{Chen2019TokenScope,Li2019,Li2018,Permenev2020,Dickerson2019,Azzopardi2019,Sun2020,Li2020,Li2019bnb,So2019,Hajdu2020Events,Viglianisi2020,Osterland2020}
and development of a complete formal specification of ERC20 in $\mathbb{K}$~\cite{ERC20K}.
Several tools focus on consistency checking between a token implementation and a
standard~\cite{Li2018,Li2020,Chen2019TokenScope}.
For example, \tool{TokenScope}~\cite{Chen2019TokenScope} identifies inconsistency in the histories
of smart contract executions,
while Li et al.~\cite{Li2020} formulate a relevant set of invariants for several standards.
Among other properties, several more tools~\cite{Viglianisi2020,Chen2020SODA,Hajdu2020Events}
check whether the contract emits the events determined by a specification, which can be
expressed in a specification language proposed by Hajdu et al.~\cite{Hajdu2020Events}.

Another common way to ensure validity of a token or other similar financial contract implementations
is through correctness conditions for functions related to token transfers, e.g.,
\texttt{transfer()}, \texttt{transferFrom()}, and \texttt{approve()} in an ERC20 contract.
For example, preconditions and postconditions of the \texttt{transfer()} function typically assert
the following statements: \emph{``the sender's balance is not less than the requested amount of tokens''}
and \emph{``the balances of both sender and receiver are updated according to the
provided amount''}~\cite{Park2018,Li2018,Viglianisi2020,Azzopardi2019,Li2019,Sun2020,Kalra2018,Liu2019a,Ahrendt2019a,Stegeman2018}.
To cover the corner cases, some of the listed specifications explicitly state that
\emph{balances of non-participating users remain unchanged}~\cite{Sun2020,Stegeman2018}.
Sun et al.~\cite{Sun2020} additionally check that \emph{the value to be transferred
is greater than zero}, a crowdfunding specified by Kalra et al.~\cite{Kalra2018}
only \emph{accepts investments bigger than a threshold limit}, while
Dickerson et al.~\cite{Dickerson2019} require \emph{user balances to be non-negative}.
The latter property can be found in specifications of banking contracts~\cite{Ahrendt2019a,Kasampalis2019},
while wallets usually require the opposite: the authors of \cite{Kalra2018,Chang2018}
request a user to \emph{respect the limit} of Ether that can be transferred out of
a contract within a transaction or a contract lifetime, respectively.
A wallet specification by Kalra et al.~\cite{Kalra2018} does not permit users
to transfer funds to themselves, while ERC20-K~\cite{ERC20K} allows self-transfers but considers
it a special case.

Formal specifications of financial contracts often contain vulnerability-related requirements,
primarily, integer overflow and underflow that may occur during a transfer~\cite{Viglianisi2020,Hajdu2019,Sun2020,Kasampalis2019,So2019,Ahrendt2019a}.
Other potential issues include event-ordering bugs: the authors of~\cite{Dickerson2019,Kolluri2019}
show how an unforeseen ordering between the invocations of \texttt{approve()}
and \texttt{transfer()} allows a malicious user to spend more funds than intended,
and how it can be prevented by an invariant.
Wang et al.~\cite{Wang2019payment} detect an extended set of
issues referred to as nondeterministic payment bugs, which include dependence manipulation
conditions. Other specifications address access control~\cite{Weiss2019,Viglianisi2020,Unal2019},
and sending funds to invalid addresses~\cite{Chen2020SODA,Chang2018}.
To assert that a crowdfunding contract can \emph{pay back all investors if the campaign is not
successful}, i.e., is not greedy, some specifications introduce temporal
properties~\cite{Sergey2018,Permenev2020}. To ensure that a smart contract has enough
funds to do so, Permenev et al~\cite{Permenev2020} require that
\textit{the escrow's balance must be at least the sum of investor
deposits unless the crowdsale is declared successful}.
Other temporal properties ensure that funds are raised only until a cap~\cite{Hajdu2019}
or a deadline is reached~\cite{Permenev2020} or specify the order of states in a crowdfunding
lifecycle and a corresponding payment procedure~\cite{Permenev2020,Suvorov2019}.

An encouraged way to ensure the correctness of financial operations is by specification of invariants over a
smart contract balance and an aggregation of user balances~\cite{Permenev2020,Hajdu2019,Li2020,Li2019bnb,Sun2020}.
The most common invariant states that \emph{the sum of user balances is equal to}
\texttt{tokenSupply}~\cite{Li2018,Li2020,Sun2020,Viglianisi2020,Permenev2020,So2019,Hajdu2019,Li2019,Shishkin2018},
where the balances of users are stored in a so-called bookkeeping mapping variable,
e.g., \texttt{balances}, and \texttt{tokenSupply} defines the total number of existing tokens.
Other token invariants establish \emph{the equivalence between sums of sender and receiver balances
before and after the function execution}~\cite{Alt2018} or require \emph{the difference between the contract balance
and the sum of all participants' bookkeeping balances to remain constant}~\cite{Wang2019b}.
These invariants can also certify correctness of inter-contract communications
of a smart contract~\cite{Permenev2020,Cecchetti2020}.
For example, an invariant proposed by Wang et al.~\cite{Wang2019VUL} ensures correctness for an
account that receives funds from a smart contract: \emph{``the amount deducted from a contract's
bookkeeping balances is always deposited into the recipient's account''}.
To facilitate specification and verification of these invariants, several tools identify
bookkeeping variables automatically~\cite{Chen2019TokenScope,Wang2019b}
or provide support for a special \emph{sum} function, that can be applied to mappings and/or
arrays~\cite{So2019,Permenev2020,Hajdu2019}.
Invariants are also used to express that the value of \texttt{tokenSupply} is immutable,
bounded, or does not decrease~\cite{Permenev2020,Bernardi2020,Li2019,Viglianisi2020,Alt2018}.
In addition, Bernardi et al.~\cite{Bernardi2020} consider invariants that apply to cryptocurrency
exchanges, such as proportional distribution of tokens: \emph{``one should not be able to exchange
an asset worth nothing with a positively valued asset''}.

Fairness in financial smart contracts can be analyzed by statistical means:
the authors of~\cite{Bigi2015} estimate the probability of money losses for
a participant of a goods exchange protocol. In a game-theoretical analysis of
a crowdfunding smart contract, Chatterjee et al.~\cite{Chatterjee2018} define the amount of
sold tokens to be an objective function. If its value can exceed the number of
available tokens, the contract is declared unfair.

The authors of~\cite{LamelaSeijas2020,Arusoaie2020} define a set of properties for financial
legal smart contracts, such as derivatives. The authors of \emph{Marlowe}~\cite{LamelaSeijas2020}
verify that all smart contracts written in this language \emph{conserve funds},
i.e., \emph{the money that comes in plus the money in the contract before the transaction
must be equal to the money that comes out plus the contract after the transaction, except in the
case of an error}. A similar idea that money are neither created nor destroyed by a smart
contract is addressed in Move~\cite{Blackshear2020} by using linear resource types for assets.
In \emph{Findel}~\cite{Arusoaie2020}, the requirements for financial derivatives include the following:
\emph{``derivatives should be free of calculation mistakes''}, \emph{``errors caused by external
sources should be handled properly''}, \emph{``accidental swaps should be avoided (i.e., the
generated transactions and contracts have the intended issuer and the intended owner)''}.

\subsubsection{Social Games}\label{sssec:properties-social}
We associate the term \emph{social game} with a class of smart contracts that regulate the
participation and interaction of multiple users, according to some predefined game rules.
Examples of such applications include
auctions~\cite{Sergey2018,Mavridou2019,Suvorov2019,Chatterjee2018,Antonino2020},
voting schemes~\cite{Antonino2020}, lotteries~\cite{Chatterjee2018}, and games, such as
gambling~\cite{Ahrendt2019a,Ellul2018,Kalra2018},
rock-paper-scissors~\cite{Sergey2018,Beckert2018,Chatterjee2018},
and Ponzi-like games~\cite{Mavridou2019}.
The payout to each participant is automatically determined by the contract source code, making the
analysis of \emph{fairness} even more relevant.
Fairness of social games can be compromised due to privacy issues which occur if participants see
each others' submissions on blockchains (C.f.~\cref{sssec:properties-privacy}).
In general, fairness specifications require user submissions to be valid and properly recorded,
after which a correctly determined winner receives his payout.
While its precise definition is subjective in the presence of several competing
users~\cite{Kalra2018}, game-theoretical techniques~\cite{Chatterjee2018,Laneve2019} analyze
utility functions of participants as a reflection of contract fairness.
The theory of mechanism design has also been applied in the verification of fairness properties of
game contracts~\cite{Liu2020TAV}.

Social games typically require each user submission to be accompanied by a participation fee, be it
a vote in a ballot~\cite{Bernardo2019}, a bid in an
auction~\cite{Sergey2018,Chatterjee2018,Antonino2020},
or a move in a game~\cite{Kalra2018} (e.g., \emph{``a minimum deposit is required to play the
game''}~\cite{Kalra2018}).
In addition to that, users should have the permission to participate (\emph{``people may vote if they are
registered voters''}~\cite{Frantz2016}), the choice submitted by a user should belong to a set of
possible values (\emph{``the vote goes for a registered candidate''}~\cite{Bernardo2019})
and, in some cases, \emph{submitted only once}~\cite{Sergey2018,Kalra2018}.
Temporal properties are often used to assure that submissions are made before a deadline
(\emph{``people must not vote after the deadline''}~\cite{Frantz2016}) or before the
results are finalized (\emph{``\texttt{bid} cannot happen after
\texttt{close}''}~\cite{Mavridou2019}).
Valid user inputs are to be correctly recorded by a smart contract~\cite{Bernardo2019,Sergey2018}.
As an example, in an auction contract specified by Sergey et al.~\cite{Sergey2018}, the
\texttt{pendingReturns} variable should store a mapping between each account and the sum of all his
transfers made to a contract.
A specification of a rock-paper-scissors contract by Sergey et al.~\cite{Sergey2018} summarizes
several of the listed requirements: \emph{``each player can only submit their non-trivial choice
once, and this choice will have to be a key from payoffMatrix in order to be recorded in the
corresponding contract field''}.
Correctness of recording in voting contracts can also be guaranteed through the use of invariants
over the number of votes~\cite{Antonino2020} and individual user weights~\cite{Li2020}.
For instance, Antonino et al.~\cite{Antonino2020} check that \emph{the number of votes initially
available is equal to the sum of votes still to be cast plus votes already cast throughout the
election process}.

After user inputs are collected, it is the responsibility of a smart contract to determine the
winner and perform the relevant transfers of cryptocurrency.
The developers of smart contracts that choose a winner randomly, such as lotteries~\cite{Atzei2019}
and casino~\cite{Kolluri2019}, should be aware of the risks introduced by dependence on external
oracles or randomization techniques, as discussed in~\cref{sssec:properties-security}.
To enforce a correct winner selection in a rock-paper-scissors contract, Beckert et
al.~\cite{Beckert2018} encode the game rules in a postcondition of the corresponding function.
The property of a simple Ponzi-scheme contract, \contract{King of the Ether Throne}~\cite{KotET},
certifies that \emph{a user should always be crowned after successfully sending Ether
to the contract}~\cite{Mavridou2019}.

Upon the outcome finalization, most of the social game contracts transfer funds to some user,
whether he is a successful gambler~\cite{Ellul2018} or an unsuccessful auction
participant~\cite{Sergey2018}.
In that sense, social games are similar to financial contracts with their properties and potential
flaws.
For example, the specification of an auction contract proposed by Sergey et al.~\cite{Sergey2018}
ensures that the smart contract does not send the funds arbitrarily and is able to pay out the
unsuccessful participants:
\emph{``an account, which is not the higher bidder, should be able to retrieve the full
amount of their bids from the contract, and do it exactly once''}.
A similar liveness property is formulated by Shishkin~\cite{Shishkin2018} for a DAO-like smart
contract, stating that \emph{if some investor deposited some amount of money and did not
vote for any proposal afterward, then he will always be able to get a refund by calling the
corresponding function}.
There are also specifications describing different ways to guarantee that a smart contract is not
greedy and has enough funds to perform the required payments.
For example, the owner of a casino in~\cite{Ellul2018} is only allowed to \emph{withdraw from the
pot when the bet of a player is resolved}.
According to Beckert et al.~\cite{Beckert2018}, \emph{as long as the auction remains
open, the sum of the funds in the auction remains the same or increases but never decreases}.

Smart contract specifications can also describe the principal rules of its execution.
For example, temporal properties that define the order of state transitions in an auction contract
are shown in~\cref{fig:auction_properties}.
Another auction specification~\cite{Beckert2019} states that \emph{the items belong to the
auctioneer as long as the auction is open and to the highest bidder after the auction is closed}.
Invariants for an auction contract in~\cite{Liu2020TAV} require that \emph{the bidder with the
highest bid becomes the winner} and \emph{the highest bid becomes the clear price}.

\subsubsection{Asset Tracking}\label{sssec:properties-resource}
Supply chain is named to be one of the most suitable use cases for permissioned
blockchains~\cite{Fournier2019}.
A supply chain typically involves several interacting parties responsible for production,
transportation, and sales.
Alqahtani et al.~\cite{Alqahtani2020} describe the responsibilities of parties in a fuel supply
chain, e.g., \emph{``vendors must always maintain their production under the daily limit and log it
into the ledger''}, \emph{``point-of-sales must always adhere to the price set by the
regulators''}, and
\emph{``once the point-of-sales receives oil, a payment should eventually be triggered''}.
Another common requirement for supply chain contracts defines the acceptable transportation
conditions, mainly, restrictions on the temperature~\cite{Fournier2019} and
humidity~\cite{Zupan2020,Wang2019Azure}.
The authors of~\cite{Wang2019Azure,Zupan2020} transfer the contract into a failed state if a
critical reading is received, while Fournier et al.~\cite{Fournier2019} report a violation if the
amount of such readings exceeds a threshold within an hour.

The next group of asset-management smart contracts includes marketplaces, often the ones that
facilitate trading between suppliers and producers of energy
resources~\cite{Nehai2018,Nehai2019,Mavridou2019}.
Similar to the properties of contracts from other domains, safety properties describe the correct
execution flow of a market by declaring the allowed and required chains of
actions~\cite{Nehai2018,Mavridou2019}, e.g., \emph{``if a user's payment has been received,
then his bill is edited before the market closure''}.
Each marketplace contract implements an algorithm that matches the received buying and selling
offers~\cite{Mavridou2019,Nehai2019,Nehai2018}.
Nehai et al.~\cite{Nehai2018} warrant a satisfactory solution by verifying that \emph{the
algorithm implements a repartition of energy proportional to payment} and that \emph{once
opened, the market will eventually be closed}.
Properties from their later proposal~\cite{Nehai2019} guarantee that the solution is optimal
in the sense that it maximizes the total number of tokens exchanged.

A common example of legal smart contracts is a license agreement~\cite{Governatori2014} for
product evaluation. Its formal specifications~\cite{Suvorov2019,Governatori2018,Hu2018}
describe the rights, obligations, and prohibitions of a user who participates in the evaluation.
For example, one of the clauses states that \emph{the Licensee must not publish comments on the
evaluation of the Product, unless the Licensee is permitted to publish the results of the
evaluation}.
As discussed in~\cref{ssec:contract-model}, logic-based languages are practical for capturing the
legal modality of such contracts~\cite{Governatori2018,Hu2018}.
Still, Suvorov et al.~\cite{Suvorov2019} encoded some of the clauses in temporal logic---although
there is no direct representation of deontic modalities,
the properties are still sufficient to identify a violation of certain contractual clauses.
This approach also enables description of common principles of a contract operation, e.g., that
\emph{removal of a comment cannot happen if nothing has been published}.
It general, formal representations of rights and obligations in smart legal contracts were
discussed with regard to various formalisms including deontic logic~\cite{Azzopardi2018},
automata-based specifications~\cite{Azzopardi2019}, a domain-specific language
based on $LDL_f$~\cite{Sato2018}, set-based frameworks~\cite{Xu2019,Banach2020}, and some others.
The authors of~\cite{Molina-Jimenez2018} verify the formal contractual model of a data selling
smart contract against typical contractual problems, such as clause duplication.

Several formal specifications exist for name registration smart contracts that manage domain names
or provide aliases to blockchain account addresses~\cite{Abdellatif2018b,Maksimov2020,Hirai2016}.
Most of them focus on security analysis~\cite{Abdellatif2018b,Hirai2016}, e.g., estimating the
chance of a stealing attack~\cite{Abdellatif2018b}.
On the contrary, Maksimov et al.~\cite{Maksimov2020} formulate a property regarding the
functional correctness of a smart contract: \emph{``the user can get an alias or a rejection after a number
of attempts not exceeding a specified value''}.

\subsection{Summary}\label{ssec:summary-verification}
Smart contract properties express security requirements and semantic conformance
in many forms. Vulnerabilities are often detected through path-level patterns
in control- and data-flow of smart contract execution.
In program-level representations, both vulnerable conditions and semantic correctness of smart
contract functions can be captured by Hoare-style properties.
The corresponding checks can as well be instrumented into the smart contract code,
which, however, increases the amount of gas consumed by a smart contract.
Nevertheless, temporal logic is considered a more suitable formalism to
specify the multi-agent execution environment of blockchain~\cite{Meyden2019}.
Indeed, contract-level behavioral specifications usually employ one of temporal
logics to formulate temporal and probabilistic constraints and account for user
strategies.
The analysis of smart contract properties outlined in~\cref{ssec:domain-properties}
also proves that the execution logic of smart contracts often relies on temporal
conditions.
However, as has been observed in~\cref{ssec:summary-forms}, contract-level modeling
and specification techniques seldom consider low-level execution details.

%% file: verification.tex
\section{Smart Contract Formal Verification Techniques}\label{sec:verification}

\begin{table}[tb]
  \begin{center}
    \caption{A (partial) summary of formal verification tools.}\label{tab:verification}
    \resizebox{.75\columnwidth}{!}{
    \begin{tabular}{L{3.8cm}L{4cm}L{6cm}r}
      \toprule
      \textbf{Verification Techniques} & \textbf{Tools} & \textbf{Selected References}
        & \textbf{Total} \\
      \midrule
      \multirow{10}{*}{Model Checking} & NuSMV, nuXmv &
      \cite{Nehai2018,Mavridou2019,Nelaturu2020,Alqahtani2020,Kongmanee2019,Madl2019,Unal2019} & %
        \multirow{10}{*}{\ModelCheckCount} \\
        & SPIN & \cite{Molina-Jimenez2018,Bai2018,Osterland2020} \\
        & CPN & \cite{Liu2019a,Duo2020} \\
        & BIP-SMC & \cite{Abdellatif2018b,Maksimov2020} \\
        & PRISM & \cite{Bigi2015} \\
        & UPPAAL & \cite{Andrychowicz2014} \\
        & MCK & \cite{Meyden2019} \\
        & Maude & \cite{Atzei2019} \\
        & FDR & \cite{Qu2018} \\
        & Ambient Calculus & \cite{Unal2019} \\
        & LDLf & \cite{Sato2018} \\
        \midrule
      \multirow{3}{*}{Theorem Proving} & Coq &
      \cite{Bernardo2019,Nielsen2019,Annenkov2020,Sergey2018,Sun2020,Yang2019a,Arusoaie2020,OConnor2017}
      & \multirow{3}{*}{\TheoremProvCount} \\
        & Isabelle/HOL & \cite{Hirai2017,Li2019,Amani2018,Genet2020,LamelaSeijas2020,Hirai2016}
        \\
        & Agda & \cite{Chapman2019} \\
       \midrule
      \multirow{9}{*}{Program Verification} & Datalog \& Souffl{\'e}
        & \cite{Brent2018,Grech2018,Perez2019,Brent2020,Tsankov2018} &
        \multirow{9}{*}{\ProgramVerCount} \\
        & Boogie Verifier \& Corral & \cite{Wang2019Azure,Antonino2020,Hajdu2019} \\
        & LLVM \& SMACK & \cite{Kalra2018,Wang2019payment} \\
        & SeaHorn & \cite{Kalra2018,Albert2019} \\
        & F* & \cite{Bhargavan2016,Grishchenko2018} \\
        & KeY & \cite{Beckert2018,Beckert2019,Ahrendt2019a} \\
        & Why3 & \cite{Nehai2019,Horta2020,Zhang2020Why3} \\
        & $\mathbb{K}$ framework & \cite{Park2018,Kasampalis2019,Jiao2020} \\
        & Custom static analyses &
        \cite{Sergey2019,LamelaSeijas2020,Feist2019,So2019,Shishkin2018,Schneidewind2020,Frank2020,Marescotti2018,Alt2018}
         \\
       \midrule
      \multirow{4}{*}{\makecell[l]{Symbolic \& \\ Concolic Execution}} & \tool{Oyente} &
      \cite{Luu2016,Torres2018,Zhou2018,Torres2019}
      & \multirow{5}{*}{\SymbolicExCount} \\
      & \tool{Mythril} & \cite{Mueller2018,Prechtel2019,Weiss2019} \\
      & \tool{teEther} & \cite{Krupp2018} \\
      & \tool{Maian} & \cite{Nikolic2018} \\
      & \tool{Manticore} & \cite{Mossberg2019} \\
        \midrule
      \multirow{6}{*}{\makecell[l]{Runtime Verification \& \\ Testing}} & \tool{ContractLarva}
      & \cite{Azzopardi2019,Ellul2018} & \multirow{6}{*}{\RuntimeVerCount} \\
      & \tool{EVM*} & \cite{Ma2019} \\
      & \tool{ReGuard} & \cite{Liu2018Reguard} \\
      & \tool{SODA} & \cite{Chen2020SODA} \\
      & \tool{Solythesis} & \cite{Li2020} \\
      & \tool{ECFChecker} & \cite{Grossman2017} \\
      \bottomrule
    \end{tabular}}
  \end{center}
\end{table}

This section discusses techniques, tools, and frameworks that are employed
for verification of smart contract models described in~\cref{sec:modeling}.
We also discuss the capabilities of these verification techniques with
respect to the properties outlined in~\cref{sec:specifications}.
A partial summary of verification techniques and employed tools is provided
in~\cref{tab:verification}.

\subsection{Model Checking}\label{ssec:model-checking}
Model checking is a well-established technique for automatically verifying a system model (with finite states) against its specification.
When applied to smart contracts, model checkers perform verification of contract-level
models, mainly transition systems, against a temporal logic
specification~\cite{Nehai2018,Mavridou2019,Alqahtani2020,Madl2019,Kongmanee2019,Abdellatif2018b,Bigi2015,Meyden2019,Maksimov2020,Atzei2019}.
We identified $\ModelCheckCount$ verification techniques that are based on model-checking, with $7$ utilizing the capabilities of \tool{NuSMV}
and \tool{nuXmv} model checkers.
The popularity of model checking is arguably caused by the suitability of both modeling and
specification formalisms to smart contracts description combined with the existence
of established automatic frameworks.
Although model checking is successful in verifying systems of several smart contracts
or users~\cite{Nelaturu2020,Madl2019,Alqahtani2020}, its limitations are induced by
the input language of a model checker and the state explosion problem~\cite{Nehai2018}.

Support of diverse transition systems and temporal properties
(C.f.~\cref{sssec:transition-systems,sssec:contract-specification})
help model checking capture different characteristics of smart contract execution,
such as concurrency~\cite{Liu2019a,Duo2020,Qu2018,Osterland2020},
nondeterminism~\cite{Abdellatif2018b,Bigi2015,Maksimov2020,Meyden2019},
or time constraints~\cite{Andrychowicz2014}.
In addition to verifying the functional correctness of one contract~\cite{Mavridou2019,Nehai2018,Kongmanee2019},
model checking handles systems of interacting smart contracts~\cite{Nelaturu2020,Alqahtani2020} and
users~\cite{Madl2019}.
Furthermore, with specifications expressed in temporal logic, model checking is able to verify liveness properties and properties of progress,
e.g., liquidity~\cite{Bartoletti2019}.
Model checkers, such as SPIN~\cite{Holzmann1997} or PAT~\cite{Sun2008pat,Liu2011pat},
additionally verify conventional requirements of concurrent systems, such as deadlock- and livelock-freedom~\cite{Bai2018,Molina-Jimenez2018}.

However, model-checking suffers from the state explosion problem, which requires the users to apply abstraction techniques for
smart contracts~\cite{Bartoletti2019,Atzei2019} or assume a set of simplifications
to its execution~\cite{Nehai2018,Kongmanee2019}.
Furthermore, Nehai et al.~\cite{Nehai2018} claim that precise modeling of the blockchain environment
is infeasible due to the limitations of the \tool{NuSMV} model-checker input language.
Indeed, as it checks a high-level representation, model-checking approaches rarely consider the details of
smart contract execution on blockchain, such as the gas mechanism or a memory model.
One of the two identified exceptions includes a schematic model of blockchain behavior,
such as mining of transactions in blocks~\cite{Abdellatif2018b}.
A more realistic model of smart contract execution is achieved by
Duo et al.~\cite{Duo2020} by combining CPN with a program logic for EVM bytecode.

\subsection{Theorem Proving}\label{ssec:theorem-proving}
Verification based on theorem proving involves encoding a system and
its desired properties into a particular mathematical logic. Then, a theorem
prover attempts to derive a formal proof of satisfaction of these properties based on the
axioms and inference rules of the formal system. Unlike model checking, which only
verifies finite-state systems, theorem proving supports verification of infinite
systems~\cite{Rushby2001}. On the other hand, theorem-proving techniques are
usually semi-automated and require human involvement and expertise.
Nevertheless, this drawback discourages neither academia nor industry from developing tools
and languages that are backed by theorem proving.
Our study identified $\TheoremProvCount$ applications of theorem proving used to derive correctness guarantees for smart contract languages, individual smart contracts, and even verification frameworks.

Coq, Isabelle/HOL, and Agda theorem provers are used to develop formal
semantics of low-~\cite{Hirai2018,Amani2018,Genet2020,Bernardo2019,OConnor2017,Chapman2019},
intermediate-~\cite{Bernardo2020albert,Li2019,Sergey2018},
and high-level~\cite{Jiao2020,Yang2018}
programming languages for smart contracts, including DSLs for financial
contracts~\cite{Arusoaie2020,LamelaSeijas2020}. However, Li et al.~\cite{Li2019}
suggest that, among them, intermediate-level languages are the most suitable for
formal verification. Annenkov et al.~\cite{Annenkov2020} note that the authors of
formal semantics often focus on meta-theory, i.e., verification of language properties,
as in case of \emph{Plutus Core}~\cite{Chapman2019} or \emph{Simplicity}~\cite{OConnor2017}.
Established through its formal semantics, general properties of \emph{Marlowe}~\cite{LamelaSeijas2020} smart contracts
include a bound on transactions that a contract can accept and preservation of money within the contract.

\tool{ConCert}~\cite{Annenkov2020} is a Coq-based framework, which allows both
meta-theoretic and functional reasoning about a (functional) language and a
smart contract, respectively.
Together with other publications~\cite{Amani2018,Bernardo2019,Yang2019a,Li2019,Arusoaie2020,LamelaSeijas2020,Nielsen2019,Annenkov2020},
it illustrates how theorem proving helps to precisely describe and prove correctness
conditions of smart contract execution.
These conditions include Hoare-style correctness properties over the state of a
smart contract and its environment~\cite{Amani2018,Li2019,Yang2019a,Bernardo2019},
security requirements~\cite{Sun2020,Arusoaie2020}, and gas consumption
reasoning~\cite{Genet2020}.

Despite the potential expressiveness of theorem-proving approaches, they seldom
consider inter-contract communication and temporal properties of smart contracts.
An attempt to examine smart contract interactions in Coq is performed by Nielsen and
Spitters~\cite{Nielsen2019}, who prove a voting contract invariant which approximates a temporal property.
Verification of temporal properties, including liveness, in \emph{Scilla} smart
contracts, is enabled by a formalization of its trace semantics~\cite{Sergey2018},
while an embedding of the language in Coq is under development~\cite{Sergey2019}.

\subsection{Symbolic Execution}\label{ssec:symbolic-execution}
Symbolic execution executes a program symbolically and thus is able to explore multiple concrete execution paths at a time. Symbolic execution of both Ethereum and EOSIO smart contracts is based on a traversal of a CFG reconstructed from the bytecode of a smart
contract~\cite{Luu2016,He2020}. Thereby, this approach inherits the difficulties associated with the
issues of precise CFG modeling in smart contracts~(C.f.~\cref{sssec:cfg}).
Other difficulties encountered in symbolic execution for smart contracts include heavy usage of hash functions (hard to be solved by SMT solvers) and the need to symbolically model memory and smart contract interactions.
Similar to model checking, symbolic execution approaches usually explore paths up
to a certain length, which calls for the application of abstract interpretation
and partial-order reduction.
Nevertheless, symbolic execution is one of the principal approaches to detect smart contract vulnerabilities:
overall, we found $\SymbolicExCount$ applications of symbolic and concolic executions. For solving path constraints,
nine-tenths of these techniques rely on an off-the-shelf constraint solver \tool{Z3},
however, Frank et al.~\cite{Frank2020} show that other solvers, especially \tool{Yices2},
significantly outperforms \tool{Z3}.
Symbolic execution is also used to guide smart fuzzing approaches~\cite{He2019}.

In most cases, symbolic execution is used to detect predefined vulnerable patterns
in the bytecode level. For example, that is performed by primary symbolic-execution
tools that include \tool{Oyente}~\cite{Luu2016}, \tool{Mythril}~\cite{Mueller2018}, and their
extensions~\cite{Torres2019,Torres2018,Zhou2018,Prechtel2019}. \tool{EOSafe}~\cite{He2020}
verifies EOSIO smart contracts against a set of predefined vulnerabilities.
\tool{teEther}~\cite{Krupp2018} examines execution paths of interest that
contain a vulnerable instruction pattern and generates exploits to confirm the vulnerabilities.
To extend verification capabilities beyond predefined patterns, \tool{SmartScopy}~\cite{Feng2019a}
and \tool{Solar}~\cite{Li2018} provide support for user-defined vulnerable patterns and constraints, respectively.

Inter-transactional verification helps to uncover smart contract vulnerabilities that
are composite, i.e., the vulnerability can only be exploited through a sequence of transactions~\cite{Brent2020}.
However, symbolic execution suffers from the path explosion problem, which makes it unsound and unable to explore all possible
sequences due to complex constraints. Therefore, most of
the techniques, e.g., \tool{Maian}~\cite{Nikolic2018}, bound the invocation depth---the length of a considered sequence of transactions.
\tool{VerX}~\cite{Permenev2020} combines symbolic execution with abstract interpretation and models gas
mechanics of Ethereum, which allows elimination of the paths that are unreachable due
to an out-of-gas exception. The tool checks reachability of assertions inserted in
the smart contract source code from the temporal properties defined by a user. \tool{VerX}
also supports verification of a bundle of interacting smart contracts, which is rarely
addressed by static analyses.

\subsection{Program Verification}\label{ssec:program-verification}
Smart contract correctness and security are verified by a wide range of program
verification techniques, which total to $\ProgramVerCount$ publications. The most numerous category
among these techniques employs existing verification languages supported by verification
frameworks, e.g., Boogie, LLVM, Datalog, or F*. Many of program verification tools are
automated~\cite{Tsankov2018,Feist2019,Schneidewind2020,So2019} and can check for composite
vulnerabilities~\cite{Brent2020} and semantic correctness of a smart contract~\cite{Wang2019Azure}.
The issues associated with program verification (especially static analysis)
include the need to precisely abstract components of smart contract execution and memory
model~\cite{Schneidewind2020,Antonino2020,Hajdu2020Memory},
while the abstraction is often unsound~\cite{Schneidewind2020}.
Although static analysis tools often do not consider the gas mechanics of smart
contract execution~\cite{Antonino2020,Schneidewind2020,Kalra2018}, others specifically focus on the estimation of
gas consumption~\cite{Albert2019Gasol,Marescotti2018}.

Translation of the smart contract source code into verification languages enables the wide
capabilities of their toolchains.
For example, data- and control flow analyses
by Datalog and its Souffl{\'e} engine successfully detect vulnerabilities in
bytecode~\cite{Brent2018,Tsankov2018,Grech2018,Brent2020} and execution traces
of Ethereum~\cite{Perez2019} smart contracts. Translation of Solidity---a high-level
language---to Boogie enables verification of Hoare-style semantic properties~\cite{Wang2019Azure,Antonino2020}
and secure compilation of annotated smart contracts~\cite{Hajdu2019}.
LLVM was shown to be a suitable IR for EVM~\cite{Wang2019payment}, Solidity~\cite{Kalra2018},
and Go~\cite{Kalra2018} implementations to check for both vulnerabilities
and semantic conformance. Bhargavan et al.~\cite{Bhargavan2016} performed embedding
of both a Solidity subset and EVM bytecode in F* to allow verification of both high-level
correctness and low-level properties, e.g., related to gas consumption.
Translation of Solidity~\cite{Nehai2019}, EVM~\cite{Zhang2020Why3}, and Michelson~\cite{Horta2020}
smart contracts to WhyML enables deductive verification of Hoare-style properties.
The authors of~\cite{Nehai2019} note that this technique facilitates the evaluation of a smart
contract in a more realistic setting compared to model checking~\cite{Nehai2018},
which suffers from state explosion. However, unlike model checking~\cite{Nehai2018},
it does not allow specification of liveness properties. Deductive verification of
smart contracts written or translated to Java w.r.t. Hoare-style properties is performed
by the \tool{KeY} framework~\cite{Beckert2018,Beckert2019,Ahrendt2019a}.
A correct-by construction deductive verifier for EVM bytecode has been generated in the
$\mathbb{K}$ framework~\cite{Park2018}.

Although several of the above-listed approaches~\cite{Tsankov2018,Kalra2018} make soundness
claims, Schneidewind et al.~\cite{Schneidewind2020} demonstrate sources of unsoundness
identified in these analyses. The authors of~\cite{Schneidewind2020} propose a provably sound Horn-clause
abstraction for EVM bytecode based on an EVM formalization~\cite{Grishchenko2018}
for reachability analysis, which enables verification of both security and functional Hoare-style properties.
Other static analyzers support reasoning about execution traces~\cite{Chen2019TokenScope} and
smart contract written in \lang{Scilla}~\cite{Sergey2019},
\lang{Solidity}~\cite{Feist2019,So2019,Shishkin2018,Alt2018}
or EVM bytecode~\cite{Schneidewind2020}, and \lang{Tezla}~\cite{Reis2020}.

Some of the publications under study actively apply program verification tools
that are specifically designed to operate over particular formalisms, such as
set-based frameworks TLA+~\cite{Xu2019} and Event-B~\cite{Banach2020,Zhu2020},
a defeasible logic engine SPINdle~\cite{Governatori2018} for logic-based programs,
Tamarin prover for verification of concurrent protocols, C-language verifiers~\cite{Albert2019}.

\subsection{Runtime Verification and Testing}
Runtime verification is a lightweight verification technique that checks the properties of a running program.
Different from the techniques described in previous sections, runtime verification is
concerned with one execution trace at a time. In the domain of smart contracts, the
term \emph{trace} often denotes a sequence of instructions executed by a blockchain
platform, while it can also refer to a sequence of function invocations or events emitted
by a smart contract.
The availability of runtime information helps dynamic verification techniques mitigate
one of the main obstacles to smart contract analysis---the need to model a complex
execution environment of blockchain.
Runtime verification usually provides a reactive defense against vulnerabilities or violations
of correctness at runtime and can potentially identify vulnerable states that could
not be reached by model checking or symbolic execution, due to the state or
path explosion.
Approaches based on runtime verification and testing, including fuzzing, account for
$\RuntimeVerCount$ publications in our dataset.

Most of the runtime verification techniques perform verification on-chain, which allows
them to reject vulnerable transactions acting as an online defense system.
The approaches are based on enhancing the source code of a blockchain
client~\cite{Grossman2017,Gao2019,Torres2019aegis,Chen2020SODA,Rodler2018}
or a smart contract itself~\cite{Akca2019,Li2020,Azzopardi2018,Azzopardi2019,Stegeman2018,BogdanichEspina2019}.
Both Go and JavaScript Ethereum clients enhanced with monitoring strategies and/or taint analysis
are able to detect predefined vulnerable patterns in the instruction
trace~\cite{Gao2019,Ma2019,Grossman2017,Rodler2018}.
\tool{{\AE}gis}~\cite{Torres2019aegis} and \tool{SODA}~\cite{Chen2020SODA}
additionally support verification of user-defined patterns, which can, for example,
indicate an inconsistency between a contract and a corresponding standard.
Still, the discussion on the false positives produced by \tool{SODA}~\cite{Chen2020SODA}
shows that although the tool can access runtime information, its completeness
is still subject to the precision of the detection patterns.
Another obstacle to wide adoption of such instrumented clients in Ethereum arises from the fact that it
would require a hard fork of all EVM clients~\cite{Li2020}.

Another group of runtime approaches prevents vulnerabilities by
inserting protection code into the source code of a smart
contract~\cite{Li2020,Azzopardi2019,Akca2019,BogdanichEspina2019,Stegeman2018}.
For example, \tool{Solythesis}~\cite{Li2020} automatically
instruments a Solidity smart contract with custom invariants. Compared to Solidity,
its specification language for invariants contains additional features,
including quantifiers and sums.
Azzopardi et al.~\cite{Azzopardi2018} instrument a smart contact with monitors
that verify the compliance between a smart contract and a legal contract.
However, the assertions generate additional gas consumption during the execution
of an instrumented smart contract~\cite{Li2020}. Another limitation of this approach
is caused by the infeasibility to formulate and verify complex temporal properties,
including liveness, using such smart contract instrumentation.
While the above-listed tools perform on-chain verification, events emitted
during the execution of a Solidity or a Hyperledger Fabric smart contract
can also be checked for compliance off-chain~\cite{Fournier2019,Molina-Jimenez2018}.

Besides runtime verification, smart contracts can be dynamically verified
via fuzzing and testing techniques to identify and exploit
vulnerabilities~\cite{Wang2019b,Liu2018Reguard,Jiang2018}.
To navigate a fuzzer towards higher path coverage, some authors combine fuzzing
with taint analysis~\cite{Zhang2020Fuzzing} and symbolic
execution~\cite{Kolluri2019,He2019,Torres2020}. Symbolic execution is also used
to guide invariant-based testing of a smart contract by a federated society of bots~\cite{Viglianisi2020}.

\subsection{Other Techniques} \label{ssec:other-verification}

Lightweight smart contract analyses include lexical and syntactical checks
on Solidity source code and its tree-like
structures~\cite{Yamashita2019,Tikhomirov2018,Lu2019}, which detect vulnerable code
patterns.

An emerging line of research is focused on the application of statistical and deep
learning-based techniques to smart contract verification.
One of the first proposals in this domain~\cite{Liu2018} identifies vulnerable smart
contracts as containing irregular token sequences w.r.t. a statistical language model
of Solidity. Tann et al.~\cite{Tann2018} use recurrent neural networks to measure the similarity
of smart contract opcode to that of vulnerable smart contracts flagged by \tool{Maian}~\cite{Nikolic2018}.
Other machine- and deep-learning analyses are also used to identify vulnerable smart
contracts~\cite{Wang2020}, detect Ponzi schemes~\cite{Chen2018Ponzi}
and code clones~\cite{Gao2019clone}, or propose code updates~\cite{Huang2019Update}.
Additionally, machine-learning techniques help Chen et al.~\cite{Chen2020} identify whether an access control
check in a smart contract function is missing.

\subsection{Summary}\label{ssec:summary-verification}
The choice of a verification technique largely depends on the type of a formal model and
specification that one aims to analyze.
Temporal properties are typically verified by a model checker w.r.t. a contract-level
model of one or several interacting smart contracts.
Vulnerable patterns in the CFG of a smart contract are identified by symbolic execution
techniques.
Program verification approaches perform (often unsound) translation of a smart contract
to a verification language, which enables data- and control-flow analyses to uncover composite vulnerabilities.
Despite a high level of complexity associated with theorem proving, it is actively used to precisely specify and verify
correctness of smart contracts (usually expressed in Hoare triples), programming languages,
and verification frameworks.
As an alternative to static analyses, a smart contract can be verified in runtime, which involves
modification of a blockchain client or a smart contract itself. This approach enables
rejection of non-compliant transactions in runtime but may introduce gas or runtime overhead.

%% file: conclusion.tex
\section{Discussions}\label{sec:discussion}
In this section, we discuss our observations on the formal verification and specification
literature of smart contracts, to formulate answers to the research questions raised
in~\cref{ssec:methodology}.
Some key observations are also summarized in~\cref{tab:summary}.

\subsection{Results}\label{ssec:results}
With \textbf{RQ1}, we aimed to identify formalisms which are currently used in the modeling and
specification of smart contracts, as well as approaches to their verification.
Through our study, we made the following observations.
(1) Among contract-level representations, the dominant combination is state-transition models with
specifications in temporal logic, which are then verified by model checkers.
As seen from~\cref{tab:summary}, this combination is mostly used to formulate functional
correctness properties, and some security properties involving the concept of progress, such as
liquidity~\cite{Bartoletti2019}, are modeled as liveness.
(2) Another popular approach is automated program-level verification against Hoare-style
specifications.
In terms of functional correctness, program logics defined at the bytecode-level allow the
reasoning about smart contract behaviors using theorem provers, while source-level annotations are
usually discharged by program verification techniques.
Verification of security properties is dominated by existing well-established symbolic execution
and program verification tool-chains, e.g., Boogie.
With a few exceptions, these approaches identify security issues along program paths that match
predefined vulnerable patterns.
(3) Runtime verification checks for similar issues but on runtime execution traces.
These techniques often require instrumentation of smart contract code with proper monitors, e.g.,
assertions.
Runtime monitoring can also be implemented off-chain with tracking of events for compliance
checking for escrow and supply chain applications~\cite{Molina-Jimenez2018,Fournier2019}.

\begin{table}[t]
    \caption{A (partial) overview of the formalization and verification
    literature.}\label{tab:summary}
    \footnotesize
    \resizebox{\columnwidth}{!}{
    \begin{tabular}{c|p{2.2cm}|p{.25in}p{.25in}p{.25in}p{.25in}|p{.25in}p{.25in}p{.72in}%
        p{.72in}|p{.25in}p{.25in}p{.25in}p{.45in}p{.25in}}
    \toprule
    \multirowthead{5}{\T{\textbf{Domains}}}
    & \multirowthead{5}{\textbf{Applications}}
    & \multicolumn{4}{c|}{\textbf{Model Formalisms}}
    & \multicolumn{4}{c|}{\textbf{Specification Formalisms}}
    & \multicolumn{5}{c}{\textbf{Verification Techniques}} \\
    \cline{3-15}
    & & \T{\makecell{Process\\Algebra}} & \T{\makecell{Transition\\System}}
    & \T{\makecell{Control-Flow\\Automata}} & \T{\makecell{Program\\Logic}}
    & \T{\makecell{Temporal\\Logics}} & \T{\makecell{Other Logics}}
    & \T{\makecell{Hoare Logic}}
    & \T{\makecell{Path-Level\\Patterns}} 
    & \T{\makecell{Model\\Checking}} & \T{\makecell{Theorem\\Proving}}
    & \T{\makecell{Symbolic\\Execution}} & \T{\makecell{Program\\Verification}}
    & \T{\makecell{Runtime\\Verification}}
    \\ \midrule
    \multirow{8}{*}{\T{Finance}}
       & \multirowcell{2}[0ex][l]{ICO / Token}
        & \Ethereum{\cite{Li2019bnb}} & \Ethereum{\cite{Mavridou2019}} &
        \Ethereum{\cite{Permenev2020}} & \Ethereum{\cite{Park2018}}
        & \Ethereum{\cite{Mavridou2019,Permenev2020}} &
        \Ethereum{\cite{Park2018}} \Ethereum{\cite{Li2019bnb}} &
        \Ethereum{\cite{Li2020,Li2019}} & \Ethereum{\cite{Chen2019TokenScope}}
        & \Ethereum{\cite{Mavridou2019}} & \Ethereum{\cite{Li2019}} &
        \Ethereum{\cite{Permenev2020}} & \Ethereum{\cite{Li2019bnb,Park2018}} &
        \Ethereum{\cite{Li2020,Chen2019TokenScope}} 
     \\ \cline{2-15}
       & \multirowcell{2}[0ex][l]{Bank} & & \Ethereum{\cite{Kongmanee2019,Duo2020}} & &
       \Ethereum{\cite{Duo2020}}
        & \Ethereum{\cite{Kongmanee2019,Duo2020}} & & \Ethereum{\cite{Ahrendt2019a}} & \Ethereum{\cite{Bhargavan2016}}
        & \Ethereum{\cite{Kongmanee2019,Duo2020}} & & & \Ethereum{\cite{Ahrendt2019a,Bhargavan2016}} & \\
        \cline{2-15}
       & Wallet & & \Ethereum{\cite{Nelaturu2020}} & & \Tezos{\cite{Bernardo2019}}
        & \Ethereum{\cite{Nelaturu2020}} & & \Tezos{\cite{Horta2020,Bernardo2019}} &
        & \Ethereum{\cite{Nelaturu2020}} & \Tezos{\cite{Bernardo2019}} & & \Tezos{\cite{Horta2020}}
        & \\\cline{2-15}
       & \multirowcell{3}[0ex][l]{Escrow / Purchase} & \Ethereum{\cite{Qu2018}} &
       \Ethereum{\cite{Duo2020}} & &
        \Ethereum{\cite{Amani2018,Duo2020,Azzopardi2018}}
        & \Ethereum{\cite{Duo2020}} & & \Ethereum{\cite{Amani2018}} & \Ethereum{\cite{Qu2018,Azzopardi2018}} 
        & \Ethereum{\cite{Qu2018,Duo2020}} & \Ethereum{\cite{Amani2018}} & & &
        \Ethereum{\cite{Azzopardi2018}} \\
        \midrule
   \multirow{3}{*}{\T{\makecell{Social\\Games}}} & Auction
        & & \Ethereum{\cite{Mavridou2019}} & & 
        & \Ethereum{\cite{Mavridou2019}} & &
        \Ethereum{\cite{Zhang2020Why3}}\Hyperledger{\cite{Beckert2019}} & &
        \Ethereum{\cite{Mavridou2019}} & & \Ethereum{\cite{Liu2020TAV}} &
        \Ethereum{\cite{Zhang2020Why3,Liu2020TAV}}\Hyperledger{\cite{Beckert2019}} &
      \\ \cline{2-15}
     & Voting & & \Ethereum{\cite{Duo2020}} & & \Ethereum{\cite{Duo2020}}
        & \Ethereum{\cite{Duo2020}} & &
        \Tezos{\cite{Bernardo2019}}\Ethereum{\cite{Antonino2020,Li2020}} & &
        \Ethereum{\cite{Duo2020}} & \Tezos{\cite{Bernardo2019}} & \Ethereum{\cite{Liu2020TAV}} &
        \Ethereum{\cite{Antonino2020,Liu2020TAV}} & \Ethereum{\cite{Li2020}} \\
        \cline{2-15}
     & Games / Gambling & & \Ethereum{\cite{Suvorov2019}} & & 
      & \Ethereum{\cite{Suvorov2019}} & & \Hyperledger{\cite{Beckert2018}} & \Ethereum{\cite{Ellul2018}} 
      & & & & \Hyperledger{\cite{Beckert2018}} & \Ethereum{\cite{Ellul2018}}
      \\ \midrule
   \multirow{4}{*}{\T{\makecell{Asset\\Tracking}}} & Supply Chain & &
          \Hyperledger{\cite{Alqahtani2020}} & &
        & \Hyperledger{\cite{Alqahtani2020}} & & \Ethereum{\cite{Wang2019Azure}}
        & \Hyperledger{\cite{Fournier2019}}
        & \Hyperledger{\cite{Alqahtani2020}} & & & \Ethereum{\cite{Wang2019Azure}}
        & \Hyperledger{\cite{Fournier2019}}
     \\ \cline{2-15}
     & Marketplace & & \Ethereum{\cite{Nehai2018}} & &
      & \Ethereum{\cite{Nehai2018}} & & \Ethereum{\cite{Nehai2019}} &
      & \Ethereum{\cite{Nehai2018}} & & & \Ethereum{\cite{Nehai2019}} & \\ \cline{2-15}
     & License Agreement & & \Ethereum{\cite{Suvorov2019}} & &
      \Other{\cite{Governatori2018}}
      & \Ethereum{\cite{Suvorov2019}} & \Other{\cite{Governatori2018}} & &
      & & & & \Other{\cite{Governatori2018}} & \\ \cline{2-15}
      & Name Registration & & \Ethereum{\cite{Abdellatif2018b}} & &
      & \Ethereum{\cite{Abdellatif2018b}} & \Ethereum{\cite{Hirai2016}} & &
      & \Ethereum{\cite{Abdellatif2018b}} & \Ethereum{\cite{Hirai2016}} & & & \\
    \midrule
  \multirow{3}{*}{\T{Protocols}} & \multirowcell{2}[0ex][l]{Timed\\Commitment} &
  \Bitcoin{\cite{Atzei2019}} &
  \Bitcoin{\cite{Andrychowicz2014}} & &
  & \Bitcoin{\cite{Atzei2019,Andrychowicz2014}} & & &
  & \Bitcoin{\cite{Atzei2019,Andrychowicz2014}} & & & & \\\cline{2-15}
 & Atomic Swap & & \Other{\cite{Meyden2019}} & & \Other{\cite{Hirai2018}}
  & \Other{\cite{Meyden2019}} & \Other{\cite{Hirai2018}} & &
  & \Other{\cite{Meyden2019}} & \Other{\cite{Hirai2018}} & & & \\
  \midrule
   \multirow{15}{*}{\T{Security}}
      & \multirowcell{2}[0ex][l]{Reentrancy} & & \Ethereum{\cite{Mavridou2019}} &
      \Ethereum{\cite{Luu2016}} &
      & \Ethereum{\cite{Mavridou2019}} & \Ethereum{\cite{Hirai2017}} \Other{\cite{Nielsen2019}} & &
      \Ethereum{\cite{Liu2018Reguard,Wang2019payment,Luu2016}}
        & \Ethereum{\cite{Mavridou2019}} & \Ethereum{\cite{Hirai2017}} \Other{\cite{Nielsen2019}} &
        \Ethereum{\cite{Luu2016}} & \Ethereum{\cite{Wang2019payment}} & \Ethereum{\cite{Liu2018Reguard}}
      \\ \cline{2-15}
     & Concurrency & \Ethereum{\cite{Qu2018}} & & \Ethereum{\cite{Kolluri2019}} &
        & & & & \Ethereum{\cite{Qu2018,Kolluri2019,Wang2019payment}}
        & \Ethereum{\cite{Qu2018}} & & \Ethereum{\cite{Kolluri2019}} & \Ethereum{\cite{Wang2019payment}} & \\ \cline{2-15}
     & \multirowcell{2}[0ex][l]{Dependence\\Manipulation} & & & \Ethereum{\cite{Luu2016}}
     \EOS{\cite{He2020}} &
      & & & & \Ethereum{\cite{Luu2016,Wang2019payment,Ma2019}} \EOS{\cite{He2020}}
      & & & \Ethereum{\cite{Luu2016}} \EOS{\cite{He2020}} & \Ethereum{\cite{Wang2019payment}} &
      \Ethereum{\cite{Ma2019}} \\ \cline{2-15} 
     & Unchecked Call & & & \Ethereum{\cite{Luu2016}} & 
      & & & & \Ethereum{\cite{Wang2019payment,Luu2016,Chen2020SODA}}
      & & & \Ethereum{\cite{Luu2016}} & \Ethereum{\cite{Wang2019payment}} & \Ethereum{\cite{Chen2020SODA}} \\
      \cline{2-15}
     & \multirowcell{2}[0ex][l]{Access Control} & \Ethereum{\cite{Li2019bnb}} & &
     \Ethereum{\cite{Permenev2020}} \EOS{\cite{He2020}} &
      & \Ethereum{\cite{Permenev2020}} & \Ethereum{\cite{Sun2020,Li2019bnb}}
      & \Ethereum{\cite{Wang2019Azure}} & \Ethereum{\cite{Brent2020,Chen2020SODA}} %
      & & \Ethereum{\cite{Sun2020}} &
      \Ethereum{\cite{Permenev2020}} \EOS{\cite{He2020}} & \Ethereum{\cite{Brent2020,Li2019bnb,Permenev2020,Wang2019Azure}} &
      \Ethereum{\cite{Chen2020SODA}} \\ \cline{2-15}
     & \multirowcell{2}[0ex][l]{Liquidity} & \Bitcoin{\cite{Bartoletti2019}} &
     \Ethereum{\cite{Mavridou2019}} &
      \Ethereum{\cite{Nikolic2018,Tsankov2018}} &
      & \Bitcoin{\cite{Bartoletti2019}} \Ethereum{\cite{Mavridou2019}} & \Other{\cite{Sergey2018}}
      & & \Ethereum{\cite{Tsankov2018,Nikolic2018}}
      & \Bitcoin{\cite{Bartoletti2019}} \Ethereum{\cite{Mavridou2019}} & \Other{\cite{Sergey2018}}
      & \Ethereum{\cite{Nikolic2018}} & \Ethereum{\cite{Tsankov2018}} & \\ \cline{2-15}
     & \multirowcell{2}[0ex][l]{Resource\\ Consumption} & & & \Ethereum{\cite{Grech2018,Chen2017}} &
      & & \Ethereum{\cite{Genet2020}} & \Ethereum{\cite{Nehai2019}} & \Ethereum{\cite{Grech2018,Chen2017}}
      & & \Ethereum{\cite{Genet2020}} & \Ethereum{\cite{Chen2017}} &
      \Ethereum{\cite{Grech2018,Nehai2019}} & \\ \cline{2-15}
     & Arithmetic & & & \Ethereum{\cite{Torres2018}} & 
      & & \Ethereum{\cite{Sun2020}} & \Ethereum{\cite{So2019}} & \Ethereum{\cite{Ma2019,Torres2018}}
      & & \Ethereum{\cite{Sun2020}} & \Ethereum{\cite{Torres2018}} &
      \Ethereum{\cite{So2019}} & \Ethereum{\cite{Ma2019}} \\ 
    \bottomrule
\end{tabular}}

{\footnotesize \Ethereum{}: Ethereum, \Bitcoin{}: Bitcoin, \Hyperledger{}:
Hyperledger Fabric, \Tezos{}: Tezos, \EOS{}: EOS, \Other{}: Other}
\end{table}

\subsection{Open Challenges}
With \textbf{RQ2} and \textbf{RQ3}, we aimed to identify current challenges and limitations in
formal verification of smart contracts, respectively.

\paragraph{Pattern-Based Verification}
As demonstrated in \cref{tab:summary}, smart contract security analysis is mostly performed at the
program level.
These techniques typically rely on \emph{vulnerable patterns} that were defined by experts, which
limits these tools to the identification of a known set of vulnerabilities~\cite{Wang2019payment}.
The generalization to larger classes of vulnerabilities can be achieved by focusing on the root
causes of bugs, such as the influence of nondeterministic factors on smart contract
execution~\cite{Wang2019payment}.
To expand the list of detectable vulnerabilities, some works propose specification languages which
allow users to define vulnerabilities~\cite{Feng2019a}, temporal properties~\cite{Permenev2020}, or
annotate \lang{Solidity} contracts with requirements, in forms of invariants~\cite{Hajdu2019,Wang2019VUL}.

\paragraph{Limited Verification of Temporal Properties}
Temporal logic is used as a universal formalism for high-level correctness specification for smart
contracts. 
In nearly all cases, temporal properties are specified over a contract-level model
of a smart contract, which typically abstract technical details of its operation.
At the same time, program-level techniques that provide better precision in terms
of the execution environment modeling rarely support specification and verification
of temporal properties.
Nevertheless, the wide use of temporal properties in smart contract specification suggests
that they should be analyzed to the precision of program-level techniques.

\paragraph{Complications in Execution Environment Modeling}
Arguably, the major difficulty of all techniques that perform static analyses lies in
the modeling of intricate aspects of smart contract execution, which also differ across platforms.
For example, a verifier for Ethereum smart contracts should precisely model the gas mechanism~\cite{Permenev2020}
and the memory model~\cite{Antonino2020,Frank2020,Hajdu2020Memory}
with its hash-based object allocation~\cite{Permenev2020}.
Gas consumption in Ethereum is determined by the executed instructions at runtime,
so gas-aware analyses have to be carried out at the bytecode-level, which makes specification of
high-level properties less straightforward~\cite{Antonino2020}.
Precise reasoning about temporal properties concerned with physical time
is also complicated by the absence of a global clock~\cite{Chatterjee2018} and the delay in transaction processing~\cite{Miller2018}.
Execution of Ethereum contracts also introduces many sources of nondeterminism, including
transaction ordering and the fallback mechanism~\cite{Wang2019payment}, only determined at runtime.
Still, comprehensive verification of functional correctness and security of smart contracts
should allow for inter-contractual reasoning~\cite{Krupp2018}, which accounts for a nondeterministic
behavior of an external callee.
In some platforms~\cite{Hyperledger}, nondeterminism can also arise from the development of smart contracts
in general-purpose languages, which can demonstrate nondeterministic behavior. However, to ensure
consensus checking between blockchain nodes, the usage of nondeterministic methods is strongly discouraged and can
be detected statically~\cite{Yamashita2019} or in combination with dynamic analysis~\cite{Spoto2020}.
Other aspects of blockchain that require modeling include the possibility of
forks~\cite{Atzei2019a,Hirai2018} and reverted transactions.

\paragraph{Limitations of Runtime Verification}
Since smart contracts are immutable once deployed, the discussed techniques are mostly to be applied in
design-time.
Even in the domain of runtime verification, many
techniques~\cite{Li2020,Azzopardi2018,Azzopardi2019,Stegeman2018,BogdanichEspina2019}
rely on the instrumentation of contracts with protecting code, which has to be performed
pre-deployment as well.
Moreover, instrumentation introduces additional gas consumption~\cite{Li2020} and increases the
size of bytecode, which is limited~\cite{Chen2020SODA}.
Runtime verification of smart contracts is also performed through the application of instrumented
Ethereum clients~\cite{Ma2019,Gao2019,Grossman2017,Chen2020SODA}.
They, however, can only operate locally---otherwise, a hard fork is required for its integration
into a platform~\cite{Li2020}.
In case the information about the execution is retrieved from the blockchain, information about
private contract variables may not be observable~\cite{Viglianisi2020}.
Furthermore, if transactions are analyzed offline, suspicious transactions cannot be reverted in
runtime~\cite{Chen2020SODA}.

\paragraph{Fragmented Standards and Ecosystems}
Despite the extensive literature, there is still not a clear path towards safe smart contract
development~\cite{Zou2019}.
This is particularly reflected by the lack of standard development tool-chains and the difficulty in
choosing the right techniques.
Arguably, the difference between the chosen formal specification and verification techniques as well as different
threat models are the key reasons for a low agreement between smart contract verification tools
demonstrated by recent studies~\cite{Perez2019,Wang2019payment}.
In addition, our observations coincide with the opinion of He et al.~\cite{He2020} that smart
contracts operating on different platforms require substantially different approaches to
specification and vulnerability detection.
Although \cref{tab:summary} demonstrates that a vast majority of approaches concentrate
on verification of Ethereum smart contracts, our study includes the work on Bitcoin~\cite{Atzei2019,Andrychowicz2014},
Tezos~\cite{Bernardo2019,Bernardo2020albert,Reis2020}, Hyperledger Fabric~\cite{Kalra2018,Alqahtani2020,Yamashita2019,Fournier2019,Beckert2019,Sato2018,Madl2019},
EOS~\cite{He2020,Quan2019,Lee2019EOS}.

\subsection{Future Directions}
We envision a few promising research directions in the area of formal specification and
verification of smart contracts.

\paragraph{Safe Languages}
One of the emerging directions is to develop \emph{safe smart contract languages} which eliminate
many security risks by design or facilitate program verification.
For example, Tezos supports three formally certified programming languages varying in platform
abstraction level~\cite{Bernardo2019,Bernardo2020albert,Reis2020}, shipped with associated
verification frameworks.
Besides formal semantics, other guarantees are derived from linear resource types in
\lang{Move}~\cite{Blackshear2020} and \lang{Flint}~\cite{Schrans2018Flint},
explicit state changes of \lang{Bamboo}~\cite{Bamboo} and
\lang{Obsidian}~\cite{Bamboo,Coblenz2019obsidian},
functional programming principles and a restricted instruction set in
\lang{Scilla}~\cite{Sergey2019},
user-centered design as in \lang{Obsidian}~\cite{Coblenz2019obsidian}, etc.
To allow automated formal verification, some languages (e.g., DAML~\cite{daml}, Pact~\cite{pact}, and TEAL~\cite{teal})
are also intentionally designed as non-Turing complete.
We also noticed an increasing number of \emph{domain-specific languages}, for example,
\lang{Marlowe}~\cite{LamelaSeijas2020,LamelaSeijas2018} and \emph{Findel}~\cite{Biryukov2017,Arusoaie2020},
which should simplify the development and verification of financial smart contracts.

\paragraph{Self-Healing Contracts}
Another prominent direction is concerned with recovering from runtime failures/violations, a.k.a.
\emph{automated repair} of contracts.
In 2017, Magazzeni et al.~\cite{Magazzeni2017a} suggested that AI planning techniques can be used
to automatically patch the misaligned traces of smart contracts.
One of the recent examples is an automated gas-aware repair technique for Ethereum smart contracts
proposed by Yu et al.~\cite{Yu2019}, which is based on a genetic mutation technique guided towards
lower gas consumption.
To ensure validity of a generated patch w.r.t. the legitimate behavior of a vulnerable smart contract,
the authors automatically construct a test suite based on its previous blockchain transactions.
The generated patches should, therefore, fix the identified vulnerability without breaking any of these previously
passing tests. The gas constraints of the Ethereum blockchain introduce an additional criterion for
the validity of a patch, which should minimize the gas consumption.
The ability of a smart contract to recover from a violation could also be relevant in the context
of smart legal contracts, e.g., to resolve disputes and conflicts between participants of a
contract~\cite{Colombo2018}.

\paragraph{Hybrid Analyses}
In the verification of smart contacts, better results can be achieved through a combination of
online and offline techniques~\cite{Chen2020SODA,Li2020}.
A combination of contract-level and program-level approaches to smart contract modeling
can additionally help mitigate the trade-off between the high-level behavioral specification and the precise
modeling of execution details. Filling in the gap between high- and low-level
analyses can enable accurate verification of interesting temporal properties.
The study of smart contract fairness is complicated by the multi-user environment of its execution
model and unknown intentions of contract designers.
Still, to analyze such high-level properties, including fairness, researchers currently apply
techniques from the domains of game theory and mechanism
design~\cite{Laneve2019,Chatterjee2018,Liu2020TAV}.

\paragraph{Collaborative Development of Standards}
Some work has been done to promote standards and best practices in the development of
smart contracts, for example, collections of design and security~\cite{Wohrer2018,Xu2018,Wohrer2018a} patterns.
Two recent publications provided taxonomies of common smart contract properties and
invariants~\cite{Permenev2020,Bernardi2020}, while the authors of~\cite{Groce2019} summarized their
observations on the performance of static and dynamic analysis tools w.r.t. different security
issues.
The adoption of formally verified libraries is another step towards safer smart contract
development.
Two recent works~\cite{Schneidewind2020,Zhang2020Why3} used formal verification to assure
the safety of the popular \texttt{SafeMath} library for Solidity.
Other contributions include creation of a benchmark and curated datasets of vulnerable smart
contracts~\cite{Durieux2019} as well as evaluation frameworks for static analyses based on
vulnerable code injection~\cite{Akca2019,Ghaleb2020Injection}.
The development of standard benchmark is facilitating advances in the ML/DL approaches to smart
contract analysis---another trend in smart contract analysis that we observed in recent
publications.
We expect the work in this direction to continue.

%% file: conclude.tex
\section{Conclusion}\label{sec:conclude}
In view of the increasing adoption of blockchain and smart contract technologies, formal
specification and verification of smart contracts has been an active topic of research in recent
years.
Within this survey, we have presented a comprehensive analysis and classification of existing
approaches to formal modeling, specification, and verification of smart contracts.
Furthermore, we outlined common properties from different smart contract domains and correlated
them with the capabilities of existing verification techniques.
Our findings suggest that a combination of contract-level models and specifications
with model checking is widely used to reason about the functional correctness of smart
contracts from all the considered domains. 
On the other hand, program-level representations are often analyzed from the perspective of
security properties, which are checked by means of symbolic execution, program verification
toolchains, or theorem proving. Security issues can also be identified in execution traces of smart contracts
by runtime verification techniques.
A holistic perspective on smart contract analysis, which was adopted in this survey, allowed us to
uncover the existing limitations to effective formal verification of smart contracts and suggest
future research directions.
We envision further advancement in the areas including safe smart contract
language design, automated repair of smart contracts, hybrid contract- and program-level approaches
to formal modeling and specification of smart contracts, and collaborative
development of standards for safe smart contract implementation.